\documentclass[
	12pt, % allowed font sizes : 10pt (default), 11pt and 12pt
	a4paper, % paper sizes: a4paper (default), a5paper
	]{article}

\usepackage{amssymb} % advanced math symbols
\usepackage{amsmath} % for equations,  math functions and operators
\usepackage{amsthm}
\usepackage{geometry} % adjust margins below with \geometry{options}
\usepackage{graphicx} % \includegraphics preferable as pdf or png
\usepackage{caption,subcaption} % captions for figures and tables
\usepackage{setspace} % adjust line spacing with \onehalfspacing, \doublespacing and \singlespacing
\usepackage{natbib} % advanced bibliography management
\usepackage{array} % visualizes matrixes 
\usepackage[pagebackref,colorlinks,linkcolor=blue!70!black,filecolor=blue!70!black,urlcolor=blue!70!black,citecolor=blue!70!black]{hyperref}  % set hyperlinks in the document and to the outside (webpages, mail-adrdresses)
\usepackage[nameinlink]{cleveref} %for clever reference

\usepackage{enumitem} %for enumerate

\usepackage{lineno} % show line numbers in the left margin
\usepackage{bm,dsfont,sidecap,comment,sgame}

\usepackage{tikz} %tikz to draw
\usetikzlibrary{arrows,positioning,calc}
\usetikzlibrary{shapes.geometric}

\newtheorem{lemma}{Lemma}

\newtheorem{proposition}{Proposition}

\newtheorem{assumption}{Assumption}
\newtheorem{set-up}{Set-up}

\newtheorem*{theorem*}{Theorem}
\newtheorem*{corollary*}{Corollary}
\newtheorem*{lemma*}{Lemma}
\newtheorem*{observation*}{Observation}
\newtheorem*{proposition*}{Proposition}
\newtheorem*{claim*}{Claim}
\newtheorem*{fact*}{Fact}
\newtheorem*{assumption*}{Assumption}
\newtheorem*{assumptionA*}{Assumption A}
\newtheorem*{set-up*}{Set-up}

\theoremstyle{definition}

\newtheorem*{definition*}{Definition}

\newtheorem*{problem*}{Problem}
\newtheorem{example}{Example}
\newtheorem*{example*}{Example}

\usepackage{apptools}
\AtAppendix{\counterwithin{lemma}{section}}
\AtAppendix{\counterwithin{proposition}{section}}

\renewenvironment{proof}[1][\proofname]{{\noindent\textbf{#1.}}}{\qed \vspace{\topsep}}
\newcolumntype{L}[1]{>{\raggedright\let\newline\\arraybackslash\hspace{0pt}}m{#1}}
\newcolumntype{C}[1]{>{\centering\let\newline\\arraybackslash\hspace{0pt}}m{#1}}
\newcolumntype{R}[1]{>{\raggedleft\let\newline\\arraybackslash\hspace{0pt}}m{#1}}

\newcommand{\bolds}{\boldsymbol{s}}
\newcommand{\boldv}{\boldsymbol{v}}

\newcommand{\boldsigma}{\boldsymbol{\sigma}}
\newcommand{\boldc}{\boldsymbol{c}}

\newcommand{\st}{\quad \text{subject to} \quad}

\begin{document}

\begin{titlepage}
\singlespacing
\title{Decentralized Attack Search and the Design of Bug Bounty Schemes\thanks{A one-page abstract of this paper appears in the Proceedings of the 16th International Symposium on Algorithmic Game Theory, SAGT 2023. We are grateful to three anonymous referees of SAGT 2023 for their valuable suggestions. We thank Kari Kostiainen, Tapas Kundu, Priel Levy, and conference participants at the Conference on Economic Design (2023), the Annual Conference on Contests (2023), the Stony Brook International Conference on Game Theory (2023), and ETH Z\"{u}rich for their comments. This research was partially supported by the Zurich Information Security and Privacy Center (ZISC). Otto T.P. Schmidt and Pio Blieske provided excellent research assistance. All errors are our own.}}

% add the authors. Multiple authors are connected with "\and"
% respect the lexiographic order!
\author{
	Hans Gersbach\\
	\normalsize Center of Economic Research \\
        \normalsize at ETH Zurich, KOF and CEPR\\ 
	\normalsize Leonhardstrasse 21\\
	\normalsize 8092 Zurich, Switzerland\\ 
	\normalsize \href{mailto:hgersbach@ethz.ch}{hgersbach@ethz.ch}
	\and
	Akaki Mamageishvili\\
	\normalsize Offchain Labs\\ 
	\normalsize Zurich, Switzerland\\ 
	\normalsize \href{mailto:amamageishvili@offchainlabs.com}{amamageishvili@offchainlabs.com}
	\and
	Fikri Pitsuwan\\
	\normalsize Center of Economic Research\\
        \normalsize at ETH Zurich\\ 
	\normalsize Leonhardstrasse 21\\
	\normalsize 8092 Zurich, Switzerland\\ 
	\normalsize \href{mailto:fpitsuwan@ethz.ch}{fpitsuwan@ethz.ch}
	}

\date{Last updated: \today}

\maketitle
% \vspace{-0.8cm}

\thispagestyle{empty}
\end{titlepage}

\pagebreak \newpage

\begin{abstract}
    \noindent Systems and blockchains often have security vulnerabilities and can be attacked by adversaries, with potentially significant negative consequences. Therefore, infrastructure providers increasingly rely on bug bounty programs, where external individuals probe the system and report any vulnerabilities (bugs) in exchange for rewards (bounty). We develop a simple contest model of bug bounty. A group of individuals of arbitrary size is invited to undertake a costly search for bugs. The individuals differ with regard to their abilities, which we capture by different costs to achieve a certain probability to find bugs if any exist. Costs are private information. We study equilibria of the contest and characterize the optimal design of bug bounty schemes. In particular, the designer can vary the size of the group of individuals invited to search, add a paid expert, insert an artificial bug with some probability, and pay multiple prizes. We obtain the following results. First, we characterize the equilibria, establishing that any equilibrium strategy must be a threshold strategy, i.e. only agents with a cost of search below some (potentially individual) threshold participate in the bug bounty scheme. Second, we provide sufficient conditions for the equilibrium to be unique and symmetric. Third, we show that even inviting an unlimited crowd does not guarantee that the bug, if it exists, is found, unless there are agents which have zero costs, or equivalently have intrinsic gains from participating in the scheme. It may even happen that having more agents in the pool of potential participants lowers the probability of finding the bug. Fourth, adding a paid expert can increase or decrease the efficiency of the bug bounty scheme. Fifth, we illustrate how adding (known) bugs is another way to increase the likelihood that unknown bugs are found. When the additional costs of paying rewards are taken into account, it can be optimal to insert a known bug only with some probability. Sixth, we demonstrate that in a model with multiple prizes, having one prize (winner-takes-all) achieves the highest probability of finding the bug. Seventh, we identify circumstances when asymmetric equilibria occur. Lastly, we discuss how our baseline model can be extended to allow for multiple bugs, multiple experts, and heterogeneity of agents with respect to cost distributions, search times, and skills.
	\\
	% !!! no additional linespaces here !!!
	\vspace{0in}
	\vspace{0in}
	\bigskip
\end{abstract}
\thispagestyle{empty}

\pagebreak \newpage

\setcounter{page}{3}

%% Line numbering options
%\linenumbers

%% Linespacing options
%\singlespacing
\onehalfspacing
%\doublespacing

%%%%%%%%%%%%%%%%%%%%%%%%%%%%%%%%%%%%%%%%%%%%%%%%%%%%%%%%%%%%%%%%%%
%%%%%%%%%%%%%%%%%%%%%%%%%%%%%%%%%%%%%%%%%%%%%%%%%%%%%%%%%%%%%%%%%%
%%%%%%%%%%%%%%%%%%%%%%%%%%%%%%%%%%%%%%%%%%%%%%%%%%%%%%%%%%%%%%%%%%
%%%%%%%%%%%%%%%%%%%%%%%%%%%%%%%%%%%%%%%%%%%%%%%%%%%%%%%%%%%%%%%%%%

\section{Introduction}

Softwares and blockchains often have security vulnerabilities and can be attacked by adversaries, with potentially significant negative social or economic consequences. One such example occurred in 2019 when a ``significant flaw'' in the intended Swiss new e-voting system was discovered. With the danger of potential vote manipulation, the Federal Council paused the development and ordered a redesign of the system \citep{federalchancellery2019}. The attack discovery was part of a public intrusion test where everyone was allowed to probe the software and report any vulnerabilities (bug) in exchange for monetary rewards (bounty).\footnote{Participants are often called \textit{ethical hackers}, \textit{white-hats}, or \textit{security researchers}.} This type of program, often called \textit{bug bounty} or \textit{crowdsourced security}, has become a major tool for detecting software vulnerability searches used by governments, tech companies, and blockchains.\footnote{The success in recent years has led the authority to systematically adopt bug bounty programs as a main measure in government cybersecurity. In a recent press release, the Federal Council of Switzerland states that ``standardised security tests are no longer sufficient to uncover hidden loopholes. Therefore, in the future, it is intended that ethical hackers will search through the Federal Administration's productive IT systems and applications for vulnerabilities as part of so-called bug bounty programmes.'' \citep{federaldepartmentoffinance2022}} Bug bounty is particularly critical for blockchain infrastructure providers, since such projects do not have dedicated security teams testing software upgrades. Once the software is deployed, there is no turning back or any legal defense mechanisms against system exploitation.\footnote{At least until the next hard fork.}

There have been comprehensive accounts on the rules of engagement of bug bounty programs \citep{laszka2018}, on the effectiveness and best practices of such programs \citep{walshe2020,malladi2020}, and on the incentives of researchers to participate in bug bounty schemes~\citep{maillart2017}.  In this paper, we offer insights on some of the dimensions of bug bounty design, using a game-theoretic model of a simple contest building on the important work of \citet{ghosh2016} and \citet{sarne2017}, where agents with different abilities decide on whether or not to exert costly effort for finding bugs.

Several salient features of bug bounty differentiate our approach from that of the standard optimal contest literature. The design objective in traditional contests is to elicit the highest effort (or sum of efforts) from the contestants. This can be done by appropriately splitting up the prize \citep{moldovanu2001}, choosing a suitable reserve effort \citep{chawla2019}, setting an entry fee \citep{taylor1995} or by developing a revelation mechanism to select a subset of contestants from a pool of candidates \citep{mercier2018}.

We focus on the simple problem of how to maximize the likelihood to find bugs when a given amount of money is available for rewards. In particular, we will focus on three design variables for bug bounty systems. How large should the crowd of agents invited to find bugs be? Should paid experts be added to the crowd of invited bug finders? Should artificial bugs be added to the software to increase participation in bug finding and to increase the likelihood that the real bug is found?

To answer these questions and other, general questions about the nature of equilibria in bug bounty schemes, we develop a simple model of crowd-sourced security. A group of individuals of arbitrary size is invited to search for a bug. Whether a bug exists is uncertain. The individuals differ with regard to their abilities to find bugs, which we capture by different costs to achieve a certain probability to find the bug if it exists. Costs are private information. The designer of the bug bounty scheme offers a prize for the individual or the set of individuals who find the bug. The designer can vary the size of the group of individuals invited to find a bug, can add a paid expert to the crowd, and can insert an artificial bug with some probability.

We obtain the following results. First, we establish that any equilibrium strategy must be a threshold strategy, i.e. only agents with a cost of search below some (potentially individual) threshold participate in the bug bounty scheme. Second, we provide sufficient conditions for the equilibrium to be unique and symmetric. Third, we show that even inviting an unlimited crowd does not guarantee that bugs are found, unless there are agents which have zero costs, or equivalently have intrinsic gains from participating in the scheme. It may even happen that having more agents in the pool of potential participants may lower the probability of finding a bug. Fourth, adding paid agents can increase the efficiency of the bug bounty scheme, although the crowd that is attracted becomes smaller. Fifth, we illustrate how adding (known) bugs is another way to increase the likelihood that unknown bugs are found. When the additional costs of paying rewards are taken into account, it can be optimal to insert a known bug only with some probability. Sixth, we demonstrate that in a model with multiple prizes having one prize (winner-takes-all) achieves the highest probability of finding a bug. Finally, we identify circumstances when asymmetric equilibria occur. 

The paper is organized as follows: In the next section, we introduce the model. In \Cref{sec:analysis}, we characterize the equilibria and derive their properties for finding bugs. In \Cref{sec:extension}, we provide extensions when experts or artificial bugs are added and when multiple prizes are awarded. We also discuss the existence and nature of asymmetric equilibria. In \Cref{sec:discuss}, we discuss our baseline assumptions and show how they can be relaxed without much technical difficulties. \Cref{sec:conclusion} concludes. The proofs can be found in the Appendix.

%%%%%%%%%%%%%%%%%%%%%%%%%%%%%%%%%%%%%%%%%%%%%%%%%%%%%%%%%%%%%%%%%%%%%%
%%%%%%%%%%%%%%%%%%%%%%%%%%%%%%%%%%%%%%%%%%%%%%%%%%%%%%%%%%%%%%%%%%%%%%
%%%%%%%%%%%%%%%%%%%%%%%%%%%%%%%%%%%%%%%%%%%%%%%%%%%%%%%%%%%%%%%%%%%%%%

% \section{Literature} \label{sec:literature}

%%%%%%%%%%%%%%%%%%%%%%%%%%%%%%%%%%%%%%%%%%%%%%%%%%%%%%%%%%%%%%%%%%%%%%
%%%%%%%%%%%%%%%%%%%%%%%%%%%%%%%%%%%%%%%%%%%%%%%%%%%%%%%%%%%%%%%%%%%%%%
%%%%%%%%%%%%%%%%%%%%%%%%%%%%%%%%%%%%%%%%%%%%%%%%%%%%%%%%%%%%%%%%%%%%%%

\section{Model} \label{sec:model}

There are $n \geq 2$ agents invited to search for a bug. Denote the set of agents by $N = \{1,\dots,n\}$ and let $\bolds = (s_1,\dots,s_n) \in \{0,1\}^n$ denote the action profile of the agents, where $s_i = 1$ if agent $i$ searches, and otherwise $s_i = 0$. If agent $i$ decides to search, $i$ finds the bug with probability $q \in (0,1]$ at a random time $t_i$, uniformly distributed over the possible search time $[0,T]$, where $T$ is the maximal time for a search. These arrival times $t_i$ are stochastically independent across agents. Otherwise, $i$ does not find the bug. For simplicity, we assume that a bug exists, but the model can be reinterpreted as a model in which a bug exists with some probability.

A search is costly. If $s_i = 1$, agent $i$ incurs a cost $c_i$ which is private information and drawn from a continuous distribution $F$ with support $[\underline{c},\overline{c}]$, $0 \leq \underline{c} < \overline{c} \leq \infty$ and a finite probability density $f$.\footnote{The model can be extended to allow for $\underline{c} < 0$.}  We consider the case of \textsl{winner-takes-all} contest where only the first agent to find the bug receives a prize $V > 0$.\footnote{We consider multiple prizes in an extension and show that the winner-takes-all contest induces the highest level of participation by the agents.} If two (or more) agents find the bug at the same time, they share the prize. Yet  since the bug finding arrival time is uniformly distributed and stochastically independent across a discrete number of agents, the probability that this happens is zero and thus this can be neglected. The assumption also implies that agents that decide to search have the same probability to win the contest.

%Throughout the paper we assume $0 \leq \underline{c} < qV < \overline{c}$ which rules out extreme constellation where none of the agents or all of the agents search regardless of their costs. 

We write $\bolds_{-i} = (s_1,\dots,s_{i-1},s_{i+1},\dots,s_n)$, and let $S = \sum_{j} s_j$ and $S_{-i} = \sum_{j \neq i} s_j$ denote the total number of agents who search and the total number of agents other than agent $i$ who search, respectively. Given the setup, the payoff of agent $i$ is given by
\begin{equation} \label{eq:payoff}
u_i(s_i, \bolds_{-i},c_i) = s_i\left( p(\bolds_{-i}) V - c_i \right),
\end{equation}
where 
\begin{equation} \label{eq:payoffprob}
p(\bolds_{-i}) \equiv q \sum_{t=0}^{S_{-i}} \binom{S_{-i}}{t} q^t(1-q)^{S_{-i} - t}\frac{1}{t+1}
\end{equation}
is the probability that agent $i$ is the first agent to find the bug conditioning on searching. Given an action profile $\bolds$, let $B(\bolds)$ be the event that the bug is found. An important variable is the probability that the bug is found $\Pr(B(\bolds)) = 1-(1-q)^S$, which depends on the total number of agents participating in the search.

A strategy profile is denoted $\boldsigma = (\sigma_1,\dots,\sigma_n)$, where a strategy $\sigma_i: [\underline{c},\overline{c}] \rightarrow \{0,1\}$ maps an agent's private information to an action. We write $\sigma$ for the symmetric strategy profile $(\sigma,\dots,\sigma)$ when there is no risk of confusion and adopt the usual notational convention for $\boldsigma(\boldc)$, $\boldsigma_{-i}$, $\boldc_{-i}$, and $\boldsigma_{-i}(\boldc_{-i})$. Given a strategy profile $\boldsigma$, the ex-ante probability that the bug is found is then $\mathbb{E} [\Pr(B(\boldsigma(\boldc)))]$.

An important class of strategies is threshold strategies. A \textit{threshold strategy} with threshold $\hat{c}$, denoted by $\sigma_{\hat{c}}$, is characterized by

\begin{equation}\sigma_{\hat{c}}(c_i) = \left\{ \begin{array}{rcl}
 1 &  \mbox{if} & c_i \leq \hat{c} \\ 
 0 & \mbox{if} & c_i > \hat{c} \\
\end{array}\right. .
\end{equation}

A threshold strategy profile is denoted $\boldsigma_{\hat{\boldc}} = (\sigma_{\hat{c}_1},\dots, \sigma_{\hat{c}_n})$ for some threshold vector $\hat{\boldc} = (\hat{c}_1,\dots, \hat{c}_n)$. The ex-ante probability that the bug is found under a threshold strategy profile is then
\begin{equation}
\mathbb{E} [\Pr(B(\boldsigma_{\hat{\boldc}}(\boldc)))] = 1 - \prod_{i} (1-qF(\hat{c}_i)).
\end{equation}
If all agents use the same threshold strategy $\sigma_{\hat{c}}$, the ex ante probability that the bug is found becomes 
\begin{equation}P(\hat{c},q,n) \equiv 1 - (1-qF(\hat{c}))^n,
\end{equation}
which we shall call the \textit{probability of success}.

A strategy profile $\boldsigma^*$ is a \textit{Bayes Nash Equilibrium} (BNE) if for all $i$, $c$, and $s_i$, \[\mathbb{E}[u_i(\sigma^*_i(c_i),\boldsigma^*_{-i}(\boldc_{-i}),c_i)|c_i = c] \geq \mathbb{E}[u_i(s_i,\boldsigma^*_{-i}(\boldc_{-i}),c_i)|c_i = c].\]

%%%%%%%%%%%%%%%%%%%%%%%%%%%%%%%%%%%%%%%%%%%%%%%%%%%%%%%%%%%%%%%%%%%%%%
%%%%%%%%%%%%%%%%%%%%%%%%%%%%%%%%%%%%%%%%%%%%%%%%%%%%%%%%%%%%%%%%%%%%%%
%%%%%%%%%%%%%%%%%%%%%%%%%%%%%%%%%%%%%%%%%%%%%%%%%%%%%%%%%%%%%%%%%%%%%%

\section{Equilibrium Analysis} \label{sec:analysis}

This section analyzes the game. We offer a characterization of the equilibrium, discuss some important comparative statics, and examine the limit behaviors of the game as the number of agents grows large.

\subsection{Equilibrium Characterization}

We proceed as follows. First, we establish that any equilibrium strategy must be a threshold strategy. Second, we show that if the threshold cost vector is interior, then it satisfies a system of indifference conditions. Third, we propose a set of conditions for the equilibrium to be unique and symmetric. Lastly, we derive a simple and intuitive fixed-point condition for the unique equilibrium.

The first result states that the equilibrium strategies are threshold strategies.

\begin{proposition} \label{prop:threshold} 
$\boldsigma^* = \boldsigma_{\boldc^*}$ for some threshold vector $\boldc^* = (c^*_1,\dots,c^*_n)$. 
\end{proposition}

Consequently, we can analyze the game as if the strategies are the thresholds, and characterizing the equilibrium strategies then boils down to characterizing the \textit{equilibrium threshold vector}, $\boldc^* = (c^*_1,\dots,c^*_n) $. Suppose further that the equilibrium threshold vector is interior, $c^*_i \in (\underline{c},\overline{c})$ for all $i$. Then, it must satisfy the following system of indifference conditions: for all $i$,
\begin{equation} \label{eq:eqm}
c^*_i = V \Psi(\boldc^*_{-i}),
\end{equation}
where the function $\Psi: [\underline{c},\overline{c}]^{n-1} \rightarrow \mathbb{R}$ is given by
\begin{equation} \label{eq:psi}
\Psi(\hat{\boldc}_{-i}) \equiv q \sum_{K \subseteq N\setminus\{i\}} \left\{ \prod_{j \in K} F(\hat{c}_j) \prod_{j \notin K} (1-F(\hat{c}_j)) \left[\sum_{t=0}^{|K|} \binom{|K|}{t} q^t(1-q)^{|K|-t} \frac{1}{t+1} \right]\right\}.
\end{equation}

Indeed, $\Psi(\hat{\boldc}_{-i})$ denotes the probability that agent $i$ will be the winner given that the other $n-1$ agents deploy threshold strategies characterized by some threshold vector $\hat{\boldc}_{-i}$.\footnote{$\Psi(\hat{\boldc}_{-i})$ is in fact the expectation over the cost distribution of $p(\bolds_{-i})$ given that other agents follow threshold strategies.} The condition in \eqref{eq:eqm} then equates the cost and the expected benefits of search for each agent, characterizing the threshold cost such that the agent is indifferent between searching and not searching for the bug. The following proposition states some important properties of $\Psi$.

\begin{proposition} \label{prop:psi}
The following holds
    \begin{itemize}
        \item[(i)] $\Psi(\hat{c}_1,\dots, \hat{c}_{i-1},\hat{c}_{i+1},\dots,\hat{c}_n) = \Psi(\hat{c}_{\pi(1)},\dots, \hat{c}_{\pi(i-1)},\hat{c}_{\pi(i+1)},\dots,\hat{c}_{\pi(n)})$ for any permutation $\pi$,
        \item[(ii)] $\partial \Psi(\hat{\boldc}_{-i})/\partial \hat{c}_j < 0$ for all $j$ and  all $\hat{\boldc}_{-i} \in [\underline{c},\overline{c}]^{n-1}$,
        \item[(iii)] $\Psi(\underline{c},\dots,\underline{c}) = q$ and $\Psi(\overline{c},\dots,\overline{c}) = \frac{1-(1-q)^n}{n}$.
    \end{itemize}
\end{proposition}
The first property says that $\Psi$ is symmetric. The identity of the agents does not matter because agents are ex ante symmetric. The second property is that $\Psi$ is strictly decreasing in all its arguments. It holds because higher thresholds adopted by other agents increase their search probability and in turn lowers agent $i$'s probability of winning the prize. To facilitate a sharper prediction, we now impose two assumptions on $\Psi$.

\begin{assumption} \label{assump:symmetric}
$|\partial \Psi(\hat{\boldc}_{-i})/ \partial \hat{c}_j| \neq 1/V$ for all $j$ and all $\hat{\boldc}_{-i} \in [\underline{c},\overline{c}]^{n-1}$.
\end{assumption}

\begin{assumption} \label{assump:interior}
$\underline{c} < V\Psi(\underline{c},\dots,\underline{c}) = qV$ and $V\frac{1-(1-q)^n}{n} = V\Psi(\overline{c},\dots,\overline{c})  < \overline{c}$.
\end{assumption}

The first assumption ensures that the equilibrium is unique. Note that since the choice of a threshold is effectively agent $i$'s strategy, the function $V\Psi(\boldc_{-i})$ can be interpreted as agent $i$'s best-response function given the thresholds chosen by the other agents. \Cref{assump:symmetric} then demands that this best-response function has a slope that is never equal to unity. This guarantees that best-response functions cross only once, resulting in a unique equilibrium. \Cref{assump:interior} restricts the parameter values to ensure that the solution to the system of indifference conditions in \eqref{eq:eqm} is interior. With these two assumptions, we now characterize the unique equilibrium of the bug bounty game. To this end, define $\Phi: [\underline{c},\overline{c}]  \times (0,1] \times \mathbb{N} \rightarrow \mathbb{R}$ by
\begin{equation} \label{eq:phi}
     \Phi(\hat{c},q,n) \equiv \frac{P(\hat{c},q,n)}{nF(\hat{c})} = \frac{1-(1-qF(\hat{c}))^n}{nF(\hat{c})}
\end{equation}
if $\hat{c} > \underline{c}$ and $\Phi(\underline{c},q,n) \equiv q$.  Indeed, $\Phi$ is the probability that agent $i$ wins given that all other agents use the same threshold strategy. In other words, $\Phi$ is the ``slice'' of $\Psi$ along the ``diagonal'', i.e. when the arguments of $\Psi$ are all the same. As defined in \eqref{eq:phi}, $\Phi$ has an intuitive interpretation in that it is the probability that the bug is found, divided by the expected number of agents who search. The reason is that if the bug is found at all, then the agents participating in the search have the same chance to obtain the reward. We obtain

\begin{proposition} \label{prop:eqm}
Under \Cref{assump:symmetric} and \Cref{assump:interior}, the unique equilibrium is $\boldsigma_{c^*}$. The equilibrium threshold $c^* \equiv c^*(V,q,n) \in (\underline{c},\overline{c})$ is the solution to
\begin{equation} \label{eq:symeqm}
c^* = V \Phi(c^*,q,n).
\end{equation}
\end{proposition}

We henceforth refer to $\boldsigma_{c^*}$ simply as the equilibrium. To ease exposition, we suppress explicit dependence of $c^*$ and $\Phi$ on $V$, $q$, and $n$ when appropriate. Condition \eqref{eq:symeqm} is a special case of \eqref{eq:eqm}. It is an indifference condition capturing the fact that in an equilibrium, an agent of type $c^*$ must be indifferent between searching and not searching. The left-hand side is the cost of the search and the right-hand side is the expected reward: $V$ times $\Phi$.

\subsection{Comparative Statics}

We now perform comparative statics of the equilibrium.  For this purpose, we first state the properties of $\Phi(c,q,n)$. The properties of $c^*(V,q,n)$ then ensue since $c^*$ is the unique fixed point of $V\Phi(c,q,n)$. We obtain the following comparative statics results for $c^*$.

\begin{proposition} \label{prop:compstat}
$\Phi(c,q,n)$ is strictly decreasing in $c$ and strictly increasing in $q$. For $c > \underline{c}$, $\Phi(c,q,n)$ is strictly decreasing in $n$. The equilibrium threshold $c^*(V,q,n)$ is
\begin{itemize}
    \item[(i)] increasing in $V$,
    \item[(ii)] increasing in $q$, and 
    \item[(iii)] decreasing in $n$.
\end{itemize}
\end{proposition} 

\begin{figure}[ht]
   \centering
        \begin{tikzpicture}[scale=0.65]

\begin{scope}
     %AXIS
     \draw[->] (-0.1,0) -- (6,0) node[right]{\small{$c$}}; 
     \draw[->] (0,-0.1) -- (0,6);

      %GRAPH
     \draw[dashed,color=gray] (0,0) -- (6,6);
     \draw[dashed,color=gray] (1,0) -- (1,6);
     \draw[dashed,color=gray] (5.3,0)  -- (5.3,6);
     \draw[dashed,color=gray] (0,1) -- (6,1);
     \draw[dashed,color=gray] (0,5.3) -- (6,5.3);

     \draw[->] (3.4,0) -- (3.4,3.2);
     \draw[->] (2.5,0) -- (2.5,2.3);

     %X-AXIS LABELS
     \draw (1,-0.1) node[below]{\footnotesize{$\underline{c}$}} -- (1,0);
     \draw (5.3,-0.1) node[below]{\footnotesize{$\overline{c}$}} -- (5.3,0);
     \draw (3.4,0) -- (3.4,-0.8) node[below]{\footnotesize{$c^*(V'')$}};
     \draw (2.5,0) -- (2.5,-0.1) node[below]{\footnotesize{$c^*(V')$}};
     
     %Y-AXIS LABELS
     \draw (-0.1,4.5) node[left]{\footnotesize{$V''q$}} -- (0.1,4.5);
     %\draw (-0.1,2.8) node[left]{\small{$V''\frac{1-(1-q)^n}{n}$}} -- (0.1,2.8);

     \draw (-0.1,3) node[left]{\footnotesize{$V'q$}} -- (0.1,3);

     %PLOT
     \draw[ultra thick,color=blue] (1,4.5) .. controls (2.8,3.5) .. (5.3,2.8) ;

     \draw[ultra thick,color=red] (1,3) .. controls (2.8,2.3) .. (5.3,1.65) ;

     %LABEL
     \node at (2.6,-2.5) {\footnotesize{(a) \textcolor{blue}{$V''\Phi(c)$} and \textcolor{red}{$V'\Phi(c)$} } };
     
\end{scope}

\begin{scope}[xshift=235, yshift=0]
     %AXIS
     \draw[->] (-0.1,0) -- (6,0) node[right]{\small{$c$}}; 
     \draw[->] (0,-0.1) -- (0,6);

      %GRAPH
     \draw[dashed,color=gray] (0,0) -- (6,6);
     \draw[dashed,color=gray] (1,0) -- (1,6);
     \draw[dashed,color=gray] (5.3,0)  -- (5.3,6);
     \draw[dashed,color=gray] (0,1) -- (6,1);
     \draw[dashed,color=gray] (0,5.3) -- (6,5.3);

     \draw[->] (3.4,0) -- (3.4,3.2);
     \draw[->] (2.5,0) -- (2.5,2.3);

     %X-AXIS LABELS
     \draw (1,-0.1) node[below]{\footnotesize{$\underline{c}$}} -- (1,0);
     \draw (5.3,-0.1) node[below]{\footnotesize{$\overline{c}$}} -- (5.3,0);
     \draw (3.4,0) -- (3.4,-0.8) node[below]{\footnotesize{$c^*(q'')$}};
     \draw (2.5,0) -- (2.5,-0.1) node[below]{\footnotesize{$c^*(q')$}};
     
     %Y-AXIS LABELS
     \draw (-0.1,4.5) node[left]{\footnotesize{$Vq''$}} -- (0.1,4.5);
     %\draw (-0.1,2.8) node[left]{\footnotesize{$V''\frac{1-(1-q)^n}{n}$}} -- (0.1,2.8);

     \draw (-0.1,3) node[left]{\footnotesize{$Vq'$}} -- (0.1,3);

     %PLOT
     \draw[ultra thick,color=blue] (1,4.5) .. controls (2.8,3.5) .. (5.3,2.8) ;

     \draw[ultra thick,color=red] (1,3) .. controls (2.8,2.3) .. (5.3,1.65) ;

    %LABEL
     \node at (2.6,-2.5) {\footnotesize{(b) \textcolor{blue}{$V\Phi(c,q'')$} and \textcolor{red}{$V\Phi(c,q')$} } };
     
\end{scope}

\begin{scope}[xshift=470, yshift=0]
     %AXIS
     \draw[->] (-0.1,0) -- (6,0) node[right]{\small{$c$}}; 
     \draw[->] (0,-0.1) -- (0,6);

      %GRAPH
     \draw[dashed,color=gray] (0,0) -- (6,6);
     \draw[dashed,color=gray] (1,0) -- (1,6);
     \draw[dashed,color=gray] (5.3,0)  -- (5.3,6);
     \draw[dashed,color=gray] (0,1) -- (6,1);
     \draw[dashed,color=gray] (0,5.3) -- (6,5.3);

     \draw[->] (3.1,0) -- (3.1,2.95);
     \draw[->] (2.35,0) -- (2.35,2.2);

     %X-AXIS LABELS
     \draw (1,-0.1) node[below]{\footnotesize{$\underline{c}$}} -- (1,0);
     \draw (5.3,-0.1) node[below]{\footnotesize{$\overline{c}$}} -- (5.3,0);
     \draw (3.1,0) -- (3.1,-0.2) -- (4.1,-0.8) node[below]{\footnotesize{$c^*(n')$}};
     \draw (2.35,0) -- (2.35,-0.1) node[below]{\footnotesize{$c^*(n'')$}};
     
     %Y-AXIS LABELS
     \draw (-0.1,4) node[left]{\footnotesize{$Vq$}} -- (0.1,4);
     %\draw (-0.1,2.8) node[left]{\footnotesize{$V''\frac{1-(1-q)^n}{n}$}} -- (0.1,2.8);

     %PLOT
     \draw[ultra thick,color=red] (1,4) .. controls (2.8,3.1) .. (5.3,2.6) ;

     \draw[ultra thick,color=blue] (1,4) .. controls (2.7,1.8) .. (5.3,1.4) ;

     %LABEL
     \node at (2.6,-2.5) {\footnotesize{(c) \textcolor{blue}{$V\Phi(c,n'')$} and \textcolor{red}{$V\Phi(c,n')$} } };
     
\end{scope}

\end{tikzpicture}
    \caption{Comparative statics of $c^*(V,q,n)$.}
    \label{fig:example1}
\end{figure}
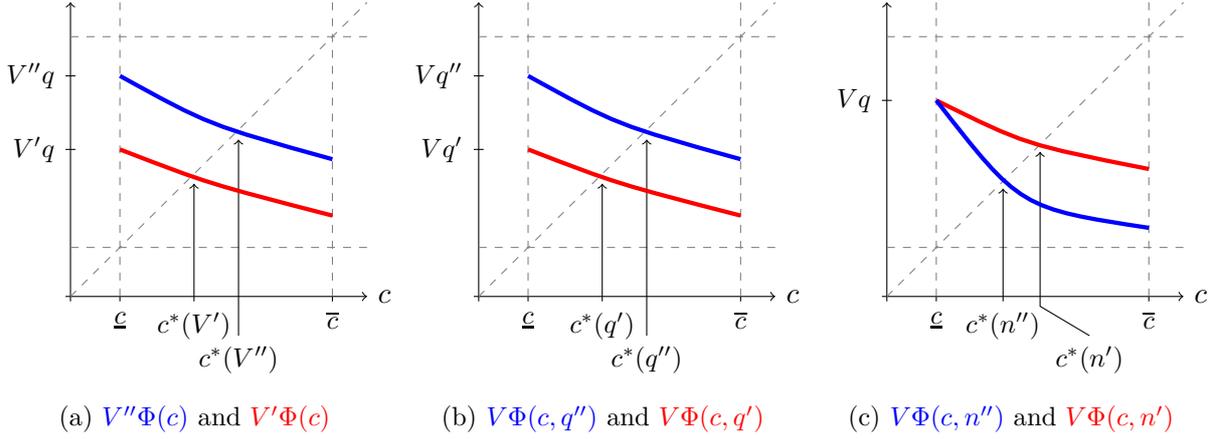

The results are intuitive. If the prize $V$ is increased, agents have more incentive to search. Agents with higher cost will now search when they otherwise would not. The same is true for when $q$, the probability that the bug is found conditioning on search, increases. Lastly, more agents intensify competition for the bug search, which lowers the probability that an agent wins the prize. \Cref{fig:example1} illustrates how $V\Phi$ changes with $V$, $q$, and $n$. Furthermore, \Cref{fig:example1} demonstrates the comparative statics of the equilibrium threshold $c^*(V,q,n)$, which is the fixed point of $V\Phi(c,q,n)$. Panel (a) of \Cref{fig:example1} shows that for $V' < V''$, $V\Phi$ as a function of $c$ shifts up with $V$, keeping $q$ and $n$ constant. Consequently, we have that $c^*(V') < c^*(V'')$. Panel (b) illustrates the case for $q' < q''$. Lastly, panel (c) illustrates that $V\Phi$ shifts down with $n$ and thus for $n' < n''$, we have $c^*(n'') < c^*(n')$.

\subsection{Probability of Success}

For the design of the bug bounty scheme, the quantity of interest is the probability of success in equilibrium, $P(c^*(V,q,n),q,n) = 1-(1-qF(c^*(V,q,n)))^n$, which we shall denote as $P^*(V,q,n)$ for simplicity.\footnote{Again, to ease exposition we suppress the arguments of $P^*$ that are kept fixed in the context of the analysis. For example, we write $c^*(n)$ and $P^*(n)$ for the equilibrium threshold and the probability of success in equilibrium, respectively, when there are $n$ agents, recognizing that $V$ and $q$ are fixed.} How does the equilibrium probability of success vary with the parameters of the model? We have the following result.

\begin{proposition} \label{prop:compstatprob} 
$P^*(V,q,n)$ increases with $V$ and $q$, and may increase or decrease with $n$.
\end{proposition} 

That $P^*$ increases with $V$ and $q$ is straightforward. The comparative statics with respect to $n$, however, is more interesting. It turns out, rather surprisingly, that the probability of success may decrease or increase with the number of agents $n$. Intuition suggests that the probability of finding the bug should go up with the number of agents. However, as we have seen, more agents result in heightened competition, which lowers the participation threshold. That is, agents crowd out each others' individual incentives to search. Either force may dominate depending on the specifications of the cost distribution and the parameters of the model. 

Since $P^*(n) = 1-(1-qF(c^*(n))^n$, there are two possible channels in which the crowding-out effect can dominate when $n$ increases. The first channel operates through the cost distribution $F$ as it can amplify a decrease in $c^*(n)$. The second channel is direct via a sharp decrease in $c^*(n)$. This happens when $c^*(n)$ starts high, perhaps due to high rewards, so that each subsequent $c^*(n)$ drops sharply relative to the increase in $n$. The following examples illustrate these two channels.

\begin{table}[t]
\centering
\vspace{0.8em}
\begin{subtable}[ht]{0.45\textwidth}
\centering
\begin{tabular}{ |c|c|c| } 
 \hline
 $n$ & $c^*(n)$ & $P^*(n)$  \\ 
 \hline \hline
 2 & 0.9151 & 0.3106 \\ 
 \hline
 3 & 0.8951 & 0.2924 \\ 
 \hline
 4 & 0.8828 & 0.2917 \\ 
 \hline
 5 & 0.8739 & 0.2948 \\ 
 \hline
 6 & 0.8669 & 0.2989 \\ 
 \hline
\end{tabular}
\caption{$F(c) = c^{20}$, $q=1$, and $V=1$.}
\label{table:example1a}
\end{subtable}
\quad
\begin{subtable}[ht]{0.45\textwidth}
\centering
\begin{tabular}{ |c|c|c| } 
 \hline
 $n$ & $c^*(n)$ & $P^*(n)$  \\ 
 \hline \hline
 2 & 0.9998 & 0.9999 \\ 
 \hline
 3 & 0.8136 & 0.9935 \\ 
 \hline
 4 & 0.7042 & 0.9923 \\ 
 \hline
 5 & 0.6301 & 0.9931 \\ 
 \hline
 6 & 0.5755 & 0.9941 \\ 
 \hline
\end{tabular}
\caption{$F(c) = c$, $q=1$, and $V = 1.999$.}
\label{table:example1b}
\end{subtable}
    \caption{$c^*(n)$ and $P^*(n)$ for \Cref{ex:probn} and \Cref{ex:probn2}.}
    \label{table:example}
\end{table}

\begin{example} \label{ex:probn}
Consider $F(c) = c^{20}$ for $0 \leq c \leq 1$, and let $q = 1$ and $V = 1$. \Cref{table:example1a} shows the numerical values of $c^*(n)$ and $P^*(n)$. The equilibrium thresholds $c^*(n)$ are decreasing in $n$ as expected. For $P^*(n)$, we see it is decreasing for $n = 2$ to $n = 4$ and increasing for $n \geq 5$ onward.
\qed
\end{example}

Intuitively, \Cref{ex:probn} demonstrates distribution functions for which most individuals are expected to have a cost close to 1, and only a few highly talented agents are expected in the pool.  Then, enlarging the pool of agents may be detrimental because as the threshold declines, the expected crowd that participates shrinks considerably making it less likely to find the bug. \Cref{ex:probn2} considers a uniform cost distribution with high rewards. Since $V$ is high, the threshold starts near 1 and declines sharply relative to the direct effect of having more agents.

\begin{example} \label{ex:probn2}
Consider $F(c) = c$ for $0 \leq c \leq 1$, and let $q = 1$ and $V = 1.999$. \Cref{table:example1b} shows the numerical values $c^*(n)$ and $P^*(n)$. The equilibrium thresholds $c^*(n)$ are decreasing in $n$ as expected. For $P^*(n)$, we see it is decreasing for $n = 2$ to $n = 4$ and increasing for $n \geq 5$ onward.
\qed
\end{example}

An implication of our analysis is that the designer of the bug bounty system should pay close attention to the number of invited agents to trade off the crowding-out effect of having many agents.

To investigate further the forces at play, we now treat $n$ as a continuous variable and calculate\footnote{A detailed derivation is provided in the proof of \Cref{prop:dpdn}.}
\begin{equation} \label{eq:dpdn}
    \frac{\mathrm{d}P^*(n)}{\mathrm{d} n} = (1-qF(c^*(n)))^n \left[ \frac{nqf(c^*(n))}{1-qF(c^*(n))} \frac{\mathrm{d}c^*(n)}{\mathrm{d}n} - \ln{(1-qF(c^*(n)))}\right].
\end{equation}
From \eqref{eq:dpdn}, we see that $\mathrm{d}P^*(n)/\mathrm{d}n \geq 0$ if and only if the magnitude of $\mathrm{d}c^*(n)/\mathrm{d}n$, which is negative by \Cref{prop:compstat}, is not too large. Using the equilibrium condition \eqref{eq:symeqm}, we can derive the following proposition.

\begin{proposition} \label{prop:dpdn}
    $\mathrm{d}P^*(n)/\mathrm{d}n \geq 0$ if and only if
\begin{equation} \label{eq:Pncondition}
\frac{(1-qF(c^*(n)))\ln{(1-qF(c^*(n)))}}{-qF(c^*(n))} \geq  \frac{1}{1 + \frac{F(c^*(n))}{c^*(n)f(c^*(n))}}.
\end{equation}
\end{proposition}

%%%%%%%%%%%%%%%%%%%%%%%%%%%%%%%%%%%%%%%%%%%%%%%%%%%%%%%%%%%%%%%%%%%%%%
%%%%%%%%%%%%%%%%%%%%%%%%%%%%%%%%%%%%%%%%%%%%%%%%%%%%%%%%%%%%%%%%%%%%%%
%%%%%%%%%%%%%%%%%%%%%%%%%%%%%%%%%%%%%%%%%%%%%%%%%%%%%%%%%%%%%%%%%%%%%%

\subsection{Large Contests} \label{sec:largecontest}

In this section, we keep all parameters fixed and examine the asymptotic behavior of the equilibrium. Throughout the section, we denote for ease of notation by $c_n = c^*(n)$ and $P_n = P^*(n)$ the equilibrium threshold and the equilibrium success probability when $n$ agents are invited to participate. Our first result asserts that the equilibrium threshold converges to $\underline{c}$.

\begin{proposition} \label{prop:clargen}
    For any $\underline{c} \geq 0$, we have $c_n \rightarrow \underline{c}$.
\end{proposition}

In other words, as the number of agents grows, individual incentives to search decrease, and in the limit only the agent with the lowest cost searches. \cite{archak2009} notes a similar result in the context of an all-pay auction. The next question of interest is the behavior of $P_n$. We show that even though the individual incentive to search decreases, the aggregate incentive goes up. 

\begin{proposition} \label{prop:largen}
The following holds:
\begin{enumerate}
    \item[(i)] If $\underline{c} = 0$, then $nF(c_n) \rightarrow \infty$. If $\underline{c} > 0$, then
    \begin{equation}
        nF(c_n) \rightarrow \kappa(\underline{c}),
    \end{equation}
    where the constant $\kappa \equiv \kappa(\underline{c})$ is the unique solution to $\underline{c} = V\frac{1-e^{-q\kappa}}{\kappa}$.

    \item[(ii)] If $\underline{c} = 0$, then $P_n \rightarrow 1$. If $\underline{c} > 0$, then
      \begin{equation}
         P_n \rightarrow 1 - e^{-q \kappa(\underline{c})}.
    \end{equation}
    \end{enumerate}
\end{proposition}

\Cref{prop:largen} has important implications for the success of bug bounty schemes. Plausibly $\underline{c} >0$ as even high-ability agents have to exert effort to find bugs. Then, even inviting an unlimited crowd to find bugs will not guarantee that bugs are found. The reason is that\textemdash given the expected intensive competition\textemdash only comparatively few agents will decide to participate and the bug is not found with some probability. Yet, if a large group of agents could be invited that are partially intrinsically motivated or by reputational concerns, cases with $\underline{c} = 0$ may become possible as well as the prospect that the bug is found with certainty.

Next, we consider the rates of convergence. We have the following result.

\begin{proposition} \label{prop:conv}
The following holds:
\begin{itemize}
    \item[(i)] $c_nF(c_n) \in \Theta(n^{-1})$,
    \item[(ii)] If $\underline{c} > 0$, then $F(c_n) \in \Theta(n^{-1})$.
\end{itemize}
\end{proposition}
A corollary of \Cref{prop:conv} is that for $F(c) = c^\alpha$ on $[0,1]$, $\alpha > 0$, $c_n \in \Theta(n^{-\frac{1}{1+\alpha}})$ and for $F(c) = \left(\frac{c-\underline{c}}{\overline{c}-\underline{c}}\right)^\alpha$ on $[\underline{c},\overline{c}]$, $\alpha > 0$, $c_n - \underline{c} \in \Theta(n^{-\frac{1}{\alpha}})$. In particular, this means that for $\mathcal{U}[0,1]$, $c_n$ converges to 0 at the rate $n^{-\frac{1}{2}}$, while for $\mathcal{U}[\underline{c},\overline{c}]$, $c_n$ converges to $\underline{c}$ at the rate $n^{-1}$.

We now investigate the tail behavior of $P_n$. In the previous section, we have shown that the probability of success may increase or decrease with the number of agents. In both examples, however, we see that $P_n$ eventually increases for large enough $n$. This is a general property as we now explore. To aid the result, we introduce an additional assumption on the cost distribution.

\begin{assumption} \label{assump:costdistribution}
    $\liminf_{c \rightarrow \underline{c}^+} \frac{F(c)}{cf(c)} = \delta$, for some $\delta > 0$.
\end{assumption}

We then have the following result. 

\begin{proposition} \label{prop:largePn}
Suppose $F$ satisfies \Cref{assump:costdistribution}. Then there exists $N$ such that for all $n > N$, $P_n$ is increasing. 
% Suppose further that the cost distribution is such that $\frac{F(c)}{cf(c)}$ is non-increasing. Then there exists $\hat{n}$ such that $P_n$ is decreasing for all $n < \hat{n}$ and increasing for all $n > \hat{n}$.
\end{proposition}

Some remarks are in order. Note that $cf(c)/F(c)$ is the elasticity of the cumulative distribution function $F$. Thus, \Cref{assump:costdistribution} says that the inverse of the elasticity of $F$ does not go to zero as $c$ approaches the lower bound of the support. In other words, we need $F$ to not change too abruptly near $\underline{c}$. \Cref{assump:costdistribution} holds for a large class of distributions. For example, for $F(c) = c^\alpha$, $\alpha > 0$ with support on $[0,1]$, we have $\frac{F(c)}{cf(c)} =  \frac{1}{\alpha} > 0$. It also holds for the Beta distribution and the exponential distribution. Lastly, \Cref{assump:costdistribution} is a sufficient condition and we conjecture that the statement that $P_n$ eventually increases holds much more generally.

%%%%%%%%%%%%%%%%%%%%%%%%%%%%%%%%%%%%%%%%%%%%%%%%%%%%%%%%%%%%%%%%%%%%%%
%%%%%%%%%%%%%%%%%%%%%%%%%%%%%%%%%%%%%%%%%%%%%%%%%%%%%%%%%%%%%%%%%%%%%%

\subsection{Uniform Cost Distribution} \label{sec:unifexample}

In this section, we consider the special case of uniform cost distribution. Given $0 \leq \underline{c} < \overline{c} < \infty$, the cumulative distribution function on the support is given by $F(c) = \frac{c-\underline{c}}{\overline{c} - \underline{c}}$. We then have
\begin{equation} \label{eq:uniformphi}
   \Phi(c,q,n) =  \left\{ \begin{array}{lcl}
\frac{\overline{c}-\underline{c}}{n(c-\underline{c})} \left[1-\left(1-\frac{q}{\overline{c}-\underline{c}}(c-\underline{c})\right)^n\right] & \mbox{for}
& \underline{c} < c \leq \overline{c} \\ q & \mbox{for} & c = \underline{c}.
\end{array}\right . 
\end{equation}

It is easy to see that $\Phi$ is strictly decreasing in $c$, strictly increasing in $q$, and strictly decreasing in $n$ on the appropriate domains. 

To illustrate our results on the limit behaviors, we now consider two numerical examples with uniform cost distribution. Let $V = 1$ and $q = 1/2$. First, consider $F \sim \mathcal{U}[0,1]$. The equilibrium threshold $c^*(n)$ solves 
\begin{equation}
(c^*(n))^2n = 1-(1-c^*(n)/2)^n.
\end{equation}
From \Cref{prop:clargen} and \Cref{prop:largen}, $c^*(n) \rightarrow 0$ and $P^*(n) \rightarrow 1$ for this distribution since $\underline{c} = 0$ and $\kappa(0) = \infty$. 

Now, consider $F \sim \mathcal{U}[1/4,5/4]$. The equilibrium threshold $c^*(n)$ solves 
\begin{equation}
nc^*(n)(c^*(n)-1/4)= 1 - (9/8-c^*(n)/2)^n.
\end{equation}
For this distribution, $c^*(n) \rightarrow \frac{1}{4}$ and $P^*(n) \rightarrow 1-e^{-\frac{1}{2}\kappa} \approx 0.797$, since $\kappa = 3.188$. \Cref{table:uniformexample} shows the numerical values of $c^*(n)$ and $P^*(n)$ for the two specifications for $n = 10, 100, 1000, 2000$.

\begin{table}[t]
\centering
\begin{subtable}[ht]{0.45\textwidth}
    \centering
\begin{tabular}{ |c|c|c| }
 \hline
 $n$ & $c^*(n)$ & $P^*(n)$  \\ 
 \hline \hline
 10 & 0.2787 & 0.7771 \\ 
 \hline
 100 & 0.0997 & 0.9939 \\ 
 \hline
 1000 & 0.0316 & 0.9999 \\
 \hline
 2000 & 0.0224 & 0.9999 \\
 \hline
\end{tabular}
\caption{$F \sim \mathcal{U}[0,1]$}
\label{table:uniformexamplea}
\end{subtable}
\quad
\begin{subtable}[ht]{0.45\textwidth}
    \centering
\begin{tabular}{ |c|c|c| }
 \hline
 $n$ & $c^*(n)$ & $P^*(n)$  \\ 
 \hline \hline
 10 & 0.3780 & 0.4839 \\ 
 \hline
 100 & 0.2767 & 0.7395 \\ 
 \hline
 1000 & 0.2531 & 0.7904 \\
 \hline
  2000 & 0.2516 & 0.7936 \\
 \hline
\end{tabular}
\caption{$F \sim \mathcal{U}[1/4,5/4]$}
\label{table:uniformexampleb}
\end{subtable}
    \caption{$c^*(n)$ and $P^*(n)$ for (a) $\mathcal{U}[0,1]$ and (b) $\mathcal{U}[1/4,5/4]$.}
    \label{table:uniformexample}
\end{table}

%%%%%%%%%%%%%%%%%%%%%%%%%%%%%%%%%%%%%%%%%%%%%%%%%%%%%%%%%%%%%%%%%%%%%%
%%%%%%%%%%%%%%%%%%%%%%%%%%%%%%%%%%%%%%%%%%%%%%%%%%%%%%%%%%%%%%%%%%%%%%

\section{Extensions} \label{sec:extension}

We provide further analysis of the bug bounty game in this section. First, we investigate how adding a non-strategic agent, interpreted as an expert, alters the equilibrium behavior. Second, we look at how adding a bug to the software can increase incentives for the agents. Third, we extend the analysis to the case of multiple prizes. Lastly, we show how asymmetric equilibria can exist without imposed assumptions.

%%%%%%%%%%%%%%%%%%%%%%%%%%%%%%%%%%%%%%%%%%%%%%%%%%%%%%%%%%%%%%%%%%%%%%
%%%%%%%%%%%%%%%%%%%%%%%%%%%%%%%%%%%%%%%%%%%%%%%%%%%%%%%%%%%%%%%%%%%%%%
%%%%%%%%%%%%%%%%%%%%%%%%%%%%%%%%%%%%%%%%%%%%%%%%%%%%%%%%%%%%%%%%%%%%%%

\subsection{Adding Experts} \label{sec:addexpert}

We next examine whether adding an expert will improve bug finding of the enlarged group\textemdash crowd plus expert. The tradeoffs are obvious. The crowd will tend to search less, but this may be overcompensated by the expert's search. Thus, suppose there is a non-strategic agent, an expert, who searches regardless of the cost and finds the bug with probability $q_e \in (0,1]$, which is common knowledge. This could arise if the bug bounty system designer outsources the search to an expert and pays for his cost. Note that we do not assume that $q_e$, in which we call \textit{expertise}, is larger than $q$. This allows us to capture the situation in which the internal security team, the ``expert'', is not necessarily more equipped to find the bugs than the crowd.\footnote{In fact, this situation is often the case in practice as \citet{malladi2020} reports: ``Systems are becoming complex, and the nature of vulnerabilities is becoming unpredictable, thereby limiting a firm’s ability to trace critical weaknesses. Given this, firms are increasingly leveraging BBPs [bug bounty programs] to crowdsource both discovery and fixing of vulnerabilities.''} We further suppose that the expert gets rewarded in the same manner as the strategic agents.\footnote{That is, the expert and the strategic agents who found the bug get rewarded with equal probability. This arises if the expert finds the bug, if any, at a random time that is also distributed uniformly on $[0,T]$. An alternative reward scheme is to keep the prize if the expert finds the bug. With this scheme, however, the equilibrium simply solves $c = V(1-q_e)\Phi(c)$.}

We now characterize the equilibrium of the game with an expert. For ease of exposition, we focus only on symmetric equilibria. Analogous to the original game (bug search without expert), the key quantity is the probability that an agent wins the prize in the game with an expert. To derive this quantity, denoted by $\Phi^e$, we condition the winning probability on two cases: if the expert does not find the bug (with probability $1-q_e$) and if the expert finds the bug (with probability $q_e$). After some algebra, we get
\begin{equation} \label{eq:phiexpert}
     \Phi^e(c,q,q_e,n) \equiv \Phi(c,q,n) - q_e \frac{1- (1-qF(c))^n(1+nqF(c))}{n(n+1)qF(c)^2}
\end{equation}
if $c > \underline{c}$ and $\Phi^e(\underline{c},q,q_e,n) \equiv q(1-q_e/2)$. Now, we assume an analog of \Cref{assump:interior} for the game with an expert to ensure the interiority of the equilibrium and obtain

\begin{proposition} \label{prop:eqmexpert}
		Suppose $\underline{c} < \Phi^e(\underline{c},q,q_e,n)$ and $\Phi^e(\overline{c},q,q_e,n) < \overline{c}$. Then, the unique symmetric equilibrium of the game with expert $q_e \in (0,1]$ is $\boldsigma_{c^e}$. The equilibrium threshold $c^e \equiv c^e(V,q,q_e,n) \in (\underline{c},\overline{c})$ is the solution to
		\begin{equation} \label{eq:eqmexpert}
		    c^e = V \Phi^e(c^e,q,q_e,n).
		\end{equation}
	 \end{proposition}

Denote $c^e(q_e)$ as the equilibrium threshold of the game with expert $q_e$ and $c^*(n)$ as the equilibrium threshold of the original game with $n$ agents. It follows that $c^e(q_e) < c^*(n)$ since the second term in \eqref{eq:phiexpert} is positive and thus $\Phi^e < \Phi$ as functions of $c$. Intuitively, the expert crowds out the search effort of the agents as fewer of them decide to search since the return prospects decline. Moreover, we have that $\lim_{q_e \rightarrow 0} c^e(q_e) = c^*(n)$ since $\Phi^e$ approaches $\Phi$ as $q_e \rightarrow 0$. This implies that for sufficiently small $q_e$, we have 
\begin{equation} \label{eq:ce1}
c^*(n+1) < c^e(q_e) < c^*(n).
\end{equation}
Furthermore, since the expert is a non-strategic agent who searches regardless of their cost, adding an expert with $q_e = q$ crowds out individuals' search incentives more so than adding an extra agent would. That is, we have that
\begin{equation} \label{eq:ce2}
    c^e(q) < c^*(n+1) < c^*(n).
\end{equation}
Together, \eqref{eq:ce1} and \eqref{eq:ce2} imply that there exists a critical expertise $\hat{q}_e \in (0,q)$ such that the equilibrium threshold in the game with an expert is equal to the equilibrium threshold in the game with an additional strategic agent, $c^e(\hat{q}_e) = c^*(n+1)$. The next proposition summarizes the above analysis.

\begin{proposition} \label{prop:cexpert}
    The critical expertise is given by $\hat{q}_e = qF(c^*(n+1))$. If $q_e < \hat{q}_e$, then $c^*(n+1) < c^e(q_e)$, while if $q_e > \hat{q}_e$, then $c^e(q_e) < c^*(n+1)$.
\end{proposition}

We now look at the probability of success when the expert is present. This probability, given by
\begin{equation}
    P^e(q_e,n) \equiv (1-q_e)P(c^e(q_e),q,n) + q_e,
\end{equation}
consists of two terms. If the expert does not find the bug then the crowd succeeds with probability $P(c^e(q_e),q,n)$, while success is guaranteed if the expert succeeds. These two terms capture the two effects. First, there is the crowding-out effect, which decreases participation and therefore decreases the probability of finding the bug. Second, there is the direct benefit of expert search, which increases the probability of finding the bug. The natural question then is whether the first or the second effect dominates, that is, whether $P^e(q_e,n)$ is larger or smaller than $P^*(n)$.

Let us first consider the extreme cases. If $q_e = 1$, then success is guaranteed as the direct benefit dominates. On the other extreme, $P^e(q_e,n) \rightarrow P^*(n)$ as $q_e \rightarrow 0$ since both effects vanish. One would then conjecture that as the expertise increases, the probability of finding the bug would also increase. It turns out that this is not the case. To see this, consider the specification from either \Cref{ex:probn} or \Cref{ex:probn2} and let $q_e = \hat{q}_e$. By \Cref{prop:cexpert} and the definition of $\hat{q}_e$, we have
\begin{equation}
\begin{aligned}
P^e(\hat{q}_e,n) &= (1-\hat{q}_e)P(c^e(\hat{q}_e),q,n) + \hat{q}_e \\ &= (1-\hat{q}_e)(1 - (1-qF(c^e(\hat{q}_e)))^n)+ \hat{q}_e \\ &= 1 - (1-\hat{q}_e)(1-qF(c^e(\hat{q}_e)))^n\\ &= P(c^*(n+1)).
\end{aligned}
\end{equation}
Therefore, the probability of success with an expert equals the probability of success with an additional strategic agent. The values from \Cref{table:example} then show that the probability of success may decrease or increase with the addition of an outside expert.

This shows that intermediate values of $q_e$ either the direct benefit or the crowding-out effect may dominate. In other words, there is non-monotonicity in the probability of success with respect to expertise. The implication is that when hiring an internal team one must make sure that their expertise is sufficiently high relative to that of the crowd.

%%%%%%%%%%%%%%%%%%%%%%%%%%%%%%%%%%%%%%%%%%%%%%%%%%%%%%%%%%%%%%%%%%%%%%
%%%%%%%%%%%%%%%%%%%%%%%%%%%%%%%%%%%%%%%%%%%%%%%%%%%%%%%%%%%%%%%%%%%%%%

\subsection{Adding Artificial Bug} \label{sec:addbug}

In this section, we allow the designer to add an artificial bug to the software, which is known to the designer, but not to the participants of the bug bounty scheme. The idea is to increase the incentives for the agents to engage in the costly search process. The downside is that the expenses of the designer are increasing as more rewards may have to be paid out.  

We assume that the event of finding the artificial bug is stochastically independent of the event of finding the real bug. This is reasonable as the designer knows nothing about the real bug. The designer selects the probability that such a known bug is found by an agent which is denoted by $q_a$. The designer can select a high (low) value by making it easy (difficult) for the artificial bug to be found by the participants in the bug bounty scheme.

We thus assume that once an agent has decided to invest in the costly search for a bug, with probability $q$ s/he finds the real bug and with probability $q_a$ s/he finds the artificial bug.  This assumption is reasonable as the search is viewed as an investment in finding the bug, which is a binary decision in our model. The equilibrium condition with a known bug is $c^* = V \Phi(c^*, q,n) + V_a \Phi(c^*,q_a,n)$, where $V_a$ is the reward to the agents if the artificial bug is found.

We observe from \Cref{prop:compstat} that $\Phi(c,q,n)$ is increasing in $q$. Hence, it is optimal for the principal to set it to set $q_a=1$ if s/he wants to maximize the probability of finding the real bug. By setting $q_a=1$,  the number of participating agents in the bug bounty scheme is maximized, and thus the chance to find the real bug. 
Hence, the optimal choice of $q_a$ corresponds to adding a very easy bug which is found by everyone with probability $1$, as long as they exert the costs to search.

Of course, adding a very easy bug will increase the expected rewards the designer has to pay to the participants. Therefore, we next look at the broader objective when the principal wants to maximize her/his utility taking into account the costs of having a real bug and the payments for rewards. Suppose the principal derives utility $W$ from finding the real bug. Then, the problem of the designer can be written as: 
\begin{equation*}
\max_{V_a,q_a} \; (W-V)P(c^*,q,n) - V_a P(c^*,q_a,n) \st c^* = V \Phi(c^*, q,n) + V_a \Phi(c^*,q_a,n).
\end{equation*}

Let us illustrate the trade-offs with a simple example. Suppose $F(c) = c$ and let $W=4$ and $V = 1$, $q = 0.5$, and $n = 2$. Without inserting a known bug, i.e. $V_a=0$ or $q_a = 0$, the equilibrium is $c^* = 4/9 \approx 0.444$ and the designer's payoff is $96/81 \approx 1.185$. With a known bug with parameters $V_a = 1$ and $q_a = 0.3$. Then, $c^* = 0.684$ and the payoff is $1.333$.

Consider the problem of inserting a known bug when $n\rightarrow \infty$. We know that $c_n$ goes to $\underline{c}$ regardless of the parameters. The asymptotic behavior of the objective function is now derived. The equilibrium condition converges to
\[\underline{c} = V \frac{1-e^{-q \kappa}}{\kappa} + V_a \frac{1-e^{-q_a\kappa}}{\kappa}\]
where $\kappa = \kappa(\underline{c},q,q_a,V,V_a)$ is increasing in both $V_a$ and $q_a$. The objective is now
\[\max_{V_a,q_a} \; (W-V)(1-e^{-q\kappa}) - V_a(1-e^{-q_a\kappa}) \st  \underline{c} = V\frac{1-e^{-q \kappa}}{\kappa} + V_a \frac{1-e^{-q_a\kappa}}{\kappa}.\]
Note that if $\underline{c} = 0$, then $\kappa = \infty$ and $V_a$ should be 0. If $\underline{c} > 0$, then there is scope for inserting a known bug as illustrated in the following example.

Suppose $F$ has support $[1/4,5/4]$ and let $W=4$ and $V = 1$, $q = 0.5$. Without inserting a known bug, i.e. $V_a=0$ or $q_a = 0$, then $\kappa = 3.188$ and the designer's payoff is $3 \cdot 0.797 = 2.390$. With a known bug with parameters $V_a = 1$ and $q_a = 0.05$. Then, $\kappa = 4.313$ and the payoff is $2.459$.

We have illustrated that there is a scope for inserting known bugs into the system. Characterizing the optimal known bug is a subject of future research.

%%%%%%%%%%%%%%%%%%%%%%%%%%%%%%%%%%%%%%%%%%%%%%%%%%%%%%%%%%%%%%%%%%%%%%
%%%%%%%%%%%%%%%%%%%%%%%%%%%%%%%%%%%%%%%%%%%%%%%%%%%%%%%%%%%%%%%%%%%%%%

\subsection{Multiple Prizes}

In this section, we extend our analysis to the case of multiple prizes. The set up is as before, but with the addition that if agent $i$ finds the bug and is the $m$-th agent to do so, agent $i$ receives a prize $v^m$ ($m = 1,\dots,n$). We denote $\boldv = (v^1,\dots,v^n)$ as the prize vector and consider $\boldv \in \mathcal{V} \equiv \{\boldv : v^1 \geq \cdots \geq v^n \geq 0 \; \text{and} \; \sum_j v^j = V \}$. The payoff of agent $i$ from~\eqref{eq:payoff} is now changed to
\begin{equation} \label{eq:multipayoff}
u_i(s_i, \bolds_{-i},c_i) = s_i\left(\sum_{m=1}^n p^m(\bolds_{-i}) v^m - c_i \right),
\end{equation}
where $p^m(\bolds_{-i})$ is now the probability that agent $i$ is the $m$-th agent to find the bug conditioning searching. The expression for $p^m(\bolds_{-i})$ is given by
\begin{equation}
p^m(\bolds_{-i}) = \left\{ \begin{array}{lcl}
     q \sum_{t=m-1}^{S_{-i}} \binom{S_{-i}}{t} q^t(1-q)^{S_{-i} - t}\frac{1}{t+1} &  \mbox{if} & m-1 \leq S_{-i} \\ 
    0 & \mbox{if} & m-1 > S_{-i}.
\end{array}\right.
\end{equation}
Note that the winner-takes-all contest is a special case with $v^1 = V$ and $p^1(\bolds_{-i}) = p(\bolds_{-i})$ as given in~\eqref{eq:payoffprob}. 

We now characterize the equilibrium of the game with the modified payoff given in~\eqref{eq:multipayoff}. We begin by nothing that \Cref{prop:threshold} still holds with essentially no modification to its proof. The equilibrium threshold vector, $\boldc^*$, if it is interior, must now satisfy the following system of indifference conditions: for all $i$, $c_i^* = \sum_{m=1}^n v^m \Psi^m(\boldc^*_{-i})$, where for $m = 1,\dots, n$, $\Psi^m: [\underline{c},\overline{c}]^{n-1} \rightarrow \mathbb{R}$ is given by
\begin{equation}
\Psi^m(\hat{\boldc}_{-i}) \equiv q \sum_{\substack{K \subseteq N\setminus\{i\} \\ |K| \geq m-1}} \left\{ \prod_{j \in K} F(\hat{c}_j) \prod_{j \notin K} (1-F(\hat{c}_j)) \left[\sum_{t=m-1}^{|K|} \binom{|K|}{t} q^t(1-q)^{|K|-t} \frac{1}{t+1} \right]\right\}.
\end{equation}
$\Psi^m$ is the probability that agent $i$ will be the $m$-th agent to find the bug given that the other $n-1$ agents deploy some threshold strategies and indeed $\Psi^1 = \Psi$. Some important properties of $\Psi^m$'s are as follows.

\begin{proposition} \label{prop:psim}
    The family of functions $\Psi^m$ ($m = 1,\dots,n$) has the following properties:
    \begin{itemize}
        \item[(i)] $\sum_{m=1}^n \Psi^m = q$,
        \item[(ii)] $\Psi^m > \Psi^{m+1}$,
        \item[(iii)] $\Psi^m$ is strictly decreasing in $c_j$ if and only if $m = 1$,
        \item[(iv)] $\Psi^1(\underline{c},\dots,\underline{c}) = q$ and $\Psi^1(\overline{c},\dots,\overline{c}) = \frac{1-(1-q)^n}{n}$, 
        \item[(v)] For $m \neq 1$, $\Psi^m(\underline{c},\dots,\underline{c}) = 0$ and $\Psi^m(\overline{c},\dots,\overline{c}) = q \sum_{t=m-1}^{n-1} \binom{n-1}{t} q^t(1-q)^{n-1-t} \frac{1}{t+1}$. 
    \end{itemize}
\end{proposition}

Some remarks are in order. First, because the agent wins some prize (not necessarily positive) with certainty if s/he finds the bug, $\sum_{m=1}^n \Psi^m = q$. Second, there is a higher probability of winning the first prize than the second. The intuition is that for a fixed number of agents who find the bug, agent $i$'s ranking is uniformly random. Given this, the first prize is always available to agent $i$ if s/he finds the bug regardless of how many other find it as well. The second prize, however, is only available if at least one other agent finds it. This reasoning leads to the fact that $\Psi^m > \Psi^{m+1}$. Third, while the probability of winning the first prize goes down as more agents participate, the probability of winning other prizes may go up. That is, $\Psi^m$ need not be strictly decreasing in $c_j$ for $m \neq 1$. To see this, consider $\Psi^2$. Intuitively, if the thresholds used by the other agents are very low, then there will be less participants and thus less agents finding the bug. In turn, this makes agent $i$'s probability of being second low as well since there is no one to be second to. Increasing the thresholds of others make them more likely to participate and find the bug, and thus increases agent $i$'s chance of being second.

As in the baseline case, we impose assumptions on $\Psi^m$ to ensure uniqueness and interiority of the equilibrium threshold, and characterize the equilibrium of the game.

\begin{proposition} \label{prop:eqmmult}
		Suppose $\sum_m v^m \partial \Psi^m/\partial c_j \neq -1$, and $\underline{c} < \sum_{m} v^m \Psi^m(\underline{c})$ and $\sum_{m} v^m \Psi^m(\overline{c}) < \overline{c}$. Then, the unique equilibrium of the game with prize vector $\boldv$ is $\boldsigma_{c^{\boldv}}$. The equilibrium threshold $c^{\boldv}$ is the solution to
		\begin{equation} \label{eq:eqmmult}
		    c^{\boldv} = \sum_{m=1}^n v^m \Phi^m(c^{\boldv}),
		\end{equation}
        where
        \begin{equation}
        \Phi^m(\hat{c}) \equiv q \sum_{k = m-1}^{n-1} \left\{ \binom{n-1}{k} F(\hat{c})^k (1-F(\hat{c}))^{n-1-k} \left[\sum_{t=m-1}^{k} \binom{k}{t} q^t(1-q)^{k-t} \frac{1}{t+1} \right]\right\}.
        \end{equation}
	 \end{proposition}

We now focus on this unique equilibrium and ask which prize allocation leads to the highest and lowest levels of participation in equilibrium. The properties from \Cref{prop:psim} and the equilibrium characterization imply the following result. 

\begin{proposition} \label{prop:optimalprize}
For any $\boldv \in \mathcal{V}$,
\begin{equation} 
V \Phi^1(\hat{c}) \geq \sum_{m=1}^n v^m \Phi^m(\hat{c}) \geq \frac{V}{n}q
\end{equation}
for all $\hat{c}$. It follows that
\begin{itemize}
    \item[(i)] The prize vector $\boldv = (V,0,\dots,0)$, i.e. the winner-takes-all contest, maximizes $c^{\boldv}$ and,  consequently, maximizes the probability of success,
    \item[(ii)] The prize vector $\boldv = (V/n,\dots,V/n)$ minimizes $c^{\boldv}$ and,  consequently, minimizes the probability of success.
\end{itemize}
\end{proposition}
A similar result has been noted in \citet{sarne2017} in a different simple contest model. Consequently, since $P^*$ is increasing in $c^*$ setting the contest to be winner-takes-all maximizes the probability of success.

Note, however, that maximizing the probability of success is typically not the principal's objective when multiple prizes are allowed. Instead, suppose the principal derives utility $W$ from finding the bug. Then, the principal's problem is to maximize
\begin{equation}
    U(\boldv) \equiv W P(c^{\boldv}) - \sum_{m=1}^n v^m P^m(c^{\boldv}),    
\end{equation}
where $P^m$ is the probability that at least $m$ agents find the bug and is given by
\begin{equation}
    P^m(c^{\boldv}) = \sum_{k = m}^n \binom{n}{k} F(c^{\boldv})^k(1-F(c^{\boldv}))^{n-k} \sum_{t=m}^k \binom{k}{t} q^t (1-q)^{k-t}.
\end{equation}
In fact, $P^1(c^{\boldv}) = P(c^{\boldv})$ and $U(\boldv)$ simplifies to $(W -v^1) P(c^{\boldv}) - \sum_{m=2}^n v^m P^m(c^{\boldv})$.

The optimal prize vector depends on the parameters of the model and is typically not the winner-takes-all structure. Characterizing the general optimal structure of the prizes is a direction of future research. Here we provide a simple example. 

\begin{example}
For $n = 2$, $v^1 + v^2 = V$ and the prize vector can be characterized by one variable $v^1$. Suppose further that $F \sim \mathcal{U}[0,1]$. The principal maximizes
\begin{equation}
    U(v^1) = 2(W-v^1)q c^{\boldv} - (W+V-2v^1)q^2 (c^{\boldv})^2.
\end{equation}
Now, the equilibrium threshold $c^{\boldv}$ solves
\begin{equation}
    c^{\boldv} = v^1 \underbrace{\left( q-\frac{q^2}{2}c^{\boldv}\right)}_{\Phi^1(c^{\boldv})} + (V-v^1) \underbrace{\frac{q^2}{2}c^{\boldv}}_{\Phi^2(c^{\boldv})} = v^1q - (2v^1-V)\frac{q^2}{2}c^{\boldv}.
\end{equation}
Combining yields
\begin{equation}
    U(v^1) = 2(W-v^1)q \frac{v^1 q}{1 + (2v^1 - V)\frac{q^2}{2}} - (W+V-2v^1)q^2 \left(\frac{v^1 q}{1 + (2v^1 - V)\frac{q^2}{2}}\right)^2.
\end{equation}
Consider $W = 2$, $V = 1$, and $q = 1$. We then have that $U(1) = 8/9 < 24/25 = U(3/4)$.
\qed
\end{example}

%%%%%%%%%%%%%%%%%%%%%%%%%%%%%%%%%%%%%%%%%%%%%%%%%%%%%%%%%%%%%%%%%%%%%%
%%%%%%%%%%%%%%%%%%%%%%%%%%%%%%%%%%%%%%%%%%%%%%%%%%%%%%%%%%%%%%%%%%%%%%

\subsection{Asymmetric Equilibria}

\Cref{prop:threshold} asserts that any equilibrium of the bug bounty game is in threshold strategies. The main analysis focuses on a symmetric equilibrium, where all agents use the same threshold. Without imposing \Cref{assump:symmetric}, however, the game may have multiple equilibria, symmetric as well as asymmetric. We illustrate this with the case of $n = 2$, where $\Psi(c_{-i}) = q[1-\frac{q}{2}F(c_{-i})]$. Assuming the equilibrium threshold vector $(c_1^*, c_2^*)$ is interior, it must solve the system of equations in \eqref{eq:eqm}. In this example, the system is
\begin{equation} \label{eq:asymeqm}
    c_1 = qV\left[1-\frac{q}{2}F(c_2)\right] \qquad \text{and} \qquad
    c_2 = qV\left[1-\frac{q}{2}F(c_1)\right].
\end{equation}

Let $q = 1$ and $V = 5/7$, and let $F$ be defined for $c \in [0,1]$ as:
\begin{equation}
    F(c) = \left\{ \begin{array}{rcl}
    \frac{14}{15}c &  \mbox{if} & 0 \leq c < \frac{3}{7} \\ 
    \frac{14}{5}c-\frac{4}{5} & \mbox{if} & \frac{3}{7} \leq c \leq \frac{4}{7} \\
    \frac{14}{30}c+\frac{8}{15} & \mbox{if} & \frac{4}{7} < c \leq 1.
\end{array}\right.
\end{equation}

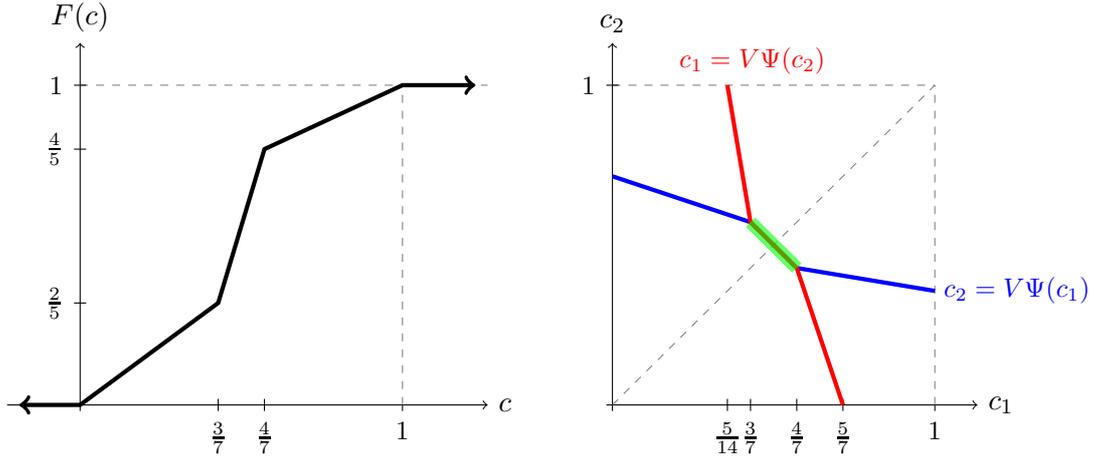
\begin{figure}[ht]
   \centering
    \begin{tikzpicture}[scale=0.8]

       \begin{scope}
     %AXIS
     \draw[->] (-1.2,0) -- (6.7,0) node[right]{\small{$c$}}; 
     \draw[->] (0,-0.1) -- (0,6) node[above]{\small{$F(c)$}};

    %GRAPH
     %\draw[dashed,color=gray] (0,0) -- (5.3,5.3);
     \draw[dashed,color=gray] (5.3,0)  -- (5.3,5.3);
     \draw[dashed,color=gray] (0,5.3) -- (6.7,5.3);

     %X-AXIS LABELS
     \draw (5.3,-0.1) node[below]{\footnotesize{$1$}} -- (5.3,0);
    %\draw (3.79,-0.1) node[below]{\footnotesize{$\frac{5}{7}$}} -- (3.79,0.1);
     \draw (3.03,-0.1) node[below]{\footnotesize{$\frac{4}{7}$}} -- (3.03,0.1);
     \draw (2.27,-0.1) node[below]{\footnotesize{$\frac{3}{7}$}} -- (2.27,0.1);
     %\draw (1.89,-0.1) node[below]{\footnotesize{$\frac{5}{14}$}} -- (1.89,0.1);

     %Y-AXIS LABELS
     \draw (-0.1,5.3) node[left]{\footnotesize{$1$}} -- (0.1,5.3);
     \draw (-0.1,1.69) node[left]{\footnotesize{$\frac{2}{5}$}} -- (0.1,1.69);
     \draw (-0.1,4.24) node[left]{\footnotesize{$\frac{4}{5}$}} -- (0.1,4.24);

     %PLOT
     \draw[ultra thick][<->] (-1,0) -- (0,0) -- (2.27,1.69) -- (3.03,4.24) -- (5.3,5.3) -- (6.5,5.3);

     %\draw[ultra thick,color=red] (1.89,5.3) -- (2.27,3.03) -- (3.03,2.27) -- (3.79,0) ;

     %LABEL
    
\end{scope}

    \begin{scope}[xshift=249,yshift = 0]
     %AXIS
     \draw[->] (-0.1,0) -- (6,0) node[right]{\small{$c_1$}}; 
     \draw[->] (0,-0.1) -- (0,6) node[above]{\small{$c_2$}};

      %GRAPH
     \draw[dashed,color=gray] (0,0) -- (5.3,5.3);
     \draw[dashed,color=gray] (5.3,0)  -- (5.3,5.3);
     \draw[dashed,color=gray] (0,5.3) -- (5.3,5.3);

     %X-AXIS LABELS
     \draw (5.3,-0.1) node[below]{\footnotesize{$1$}} -- (5.3,0);
    \draw (3.79,-0.1) node[below]{\footnotesize{$\frac{5}{7}$}} -- (3.79,0.1);
     \draw (3.03,-0.1) node[below]{\footnotesize{$\frac{4}{7}$}} -- (3.03,0.1);
     \draw (2.27,-0.1) node[below]{\footnotesize{$\frac{3}{7}$}} -- (2.27,0.1);
     \draw (1.89,-0.1) node[below]{\footnotesize{$\frac{5}{14}$}} -- (1.89,0.1);

     %Y-AXIS LABELS
     \draw (-0.1,5.3) node[left]{\footnotesize{$1$}} -- (0.1,5.3);

     %PLOT
     \draw[ultra thick,color=blue] (0,3.79) -- (2.27,3.03) -- (3.03,2.27) -- (5.3,1.89);

     \draw[ultra thick,color=red] (1.89,5.3) -- (2.27,3.03) -- (3.03,2.27) -- (3.79,0) ;

     %LABEL
     \node at (6.65,1.9) {\footnotesize{\textcolor{blue}{$c_2 = V \Psi(c_1)$} }};
      \node at (2.3,5.7) {\footnotesize{\textcolor{red}{$c_1 = V \Psi(c_2)$} }};

     \draw[line width = 5pt, green, opacity = 0.6] (2.27,3.03) -- (3.03,2.27);
\end{scope}
     
\end{tikzpicture}
    \caption{Asymmetric Equilibria Example.}
    \label{fig:asymeqm}
\end{figure}

The system in \eqref{eq:asymeqm} is depicted in \Cref{fig:asymeqm}, which shows that any 
\begin{equation} (c_1^*,c_2^*) \in \left\{(c_1,c_2): c_1 \in \left[\frac{3}{7}, \frac{4}{7}\right] \; \text{and} \; c_1 + c_2 = 1 \right\}
\end{equation}
constitutes an equilibrium threshold vector. Indeed, the symmetric equilibrium threshold $c^* = 1/2$ is one of the solutions.

%%%%%%%%%%%%%%%%%%%%%%%%%%%%%%%%%%%%%%%%%%%%%%%%%%%%%%%%%%%%%%%%%%%%%%
%%%%%%%%%%%%%%%%%%%%%%%%%%%%%%%%%%%%%%%%%%%%%%%%%%%%%%%%%%%%%%%%%%%%%%

\section{Discussions} \label{sec:discuss}

Several assumptions in the baseline model of can be relaxed without conceptual difficulty. We discuss these assumptions and show how to relax them in this section.

\subsection{Cost Distributions}

We assume that agents have the same distribution for costs in the baseline model as this is natural when agents do not know the identity of others. Since we impose very minimal assumptions on $F$, we think that a broad range of applications is covered. The distribution can be unbounded, since having $\underline{c} < 0$ does not change the results much. For instance, the limit result becomes $c_n \rightarrow \max\{0,\underline{c}\}$. The distribution could also be bimodal, perhaps modeling two pools of population: $F = aF_1 + (1-a)F_2$.

Further, relaxing the common cost distribution assumption would make the threshold values different, even when equilibrium is unique. Suppose $n=2$ and the cost distribution of agent 2 first-order stochastically dominates the cost distribution of agent 1, i.e. $F_1 \geq F_2$. That is, agent 1 is more likely to have a lower cost. Then, we expect $c_1^* > c_2^*$ because since agent 2 thinks that agent 1 is more likely to have a low cost, agent 2 would be more conservative in expending effort. To see this, let $F_1(c) = c^\alpha$ and $F_2(c) = c^\beta$ on $[0,1]$ with $\alpha < \beta$, so $c^\alpha > c^\beta$. The threshold values $c_1^*$ and $c_2^*$ solve
\[c_1 = qV\left[1-(c_2)^\beta \frac{q}{2}\right] \quad \text{and} \quad c_2 = qV\left[1-(c_1)^\alpha \frac{q}{2} \right].\]
It can be verified graphically in the $(c_1,c_2)$-space that the equations cross below the diagonal: $c_1^* > c_2^*$.

\subsection{Search Times}

Our model is a static in the sense that agents decide once whether or not to search and the prize-sharing scheme is that if more than one agent succeeds, the prize is given uniformly randomly among those who succeeded. The assumption that the search time distribution is common lays a (micro-)foundation for such prize-sharing scheme. However, this can be relaxed as follows.

Suppose search times are distributed differently, then a ``faster'' agent would have a higher threshold, since s/he can afford a more costly search effort, knowing that s/he is likely to be the first to succeed. To see this, suppose $n=2$ and let $t_1 \sim \mathcal{U}[0,T_1]$ and $t_2 \sim \mathcal{U}[0,T_2]$, with $T_1 < T_2$. This means that agent 1 is the faster agent. Then we have that the threshold values $c_1^*$ and $c_2^*$ solve
\[c_1 = qV\left[1-qF(c_2)\frac{T_1}{2T_2}\right] \quad \text{and} \quad c_2 = qV\left[1-qF(c_1)\left(1-\frac{T_1}{2T_2}\right)\right].\]
Letting $F(c) = c$, it can be easily seen graphically that $c_1^* > c_2^*$. The same conclusion should hold for other cost distributions. Moreover, the intuition extends to the case when one search time distribution first-order stochastically dominates another. In this case, the agent with a stochastically dominant search time (slower) has a lower equilibrium threshold.

\subsection{Skills}

Third, we can also extend the model to incorporate heterogeneous skills $q_i$'s that are common knowledge. The equilibrium strategies would still be in threshold strategy, but the threshold values would now solve a system of the form: 
\[c^*_i = q_i V \Xi(c^*_{-i}, q_{-i}),\]
where $\Xi$ is the analogue of $\Psi/q$ when skills are heterogeneous. Again, the threshold values will change. For two agents with $q_1 > q_2$, we can show that $c_1^* > c_2^*$ if $F$ is such that the equilibrium thresholds $c_1^*$ and $c^*_2$ are unique.

%%%%%%%%%%%%%%%%%%%%%%%%%%%%%%%%%%%%%%%%%%%%%%%%%%%%%%%%%%%%%%%%%%%%%%
%%%%%%%%%%%%%%%%%%%%%%%%%%%%%%%%%%%%%%%%%%%%%%%%%%%%%%%%%%%%%%%%%%%%%%

\subsection{Multiple Experts}

\Cref{sec:addexpert} discusses the effect of adding an expert. Having multiple experts does not qualitatively change the result. This is because an expert is modeled as a non-strategic (always exert effort) agent with a different skill $q_e$.  Adding an expert shifts the equilibrium threshold by the same amount as adding a (fractional) player. Therefore, the model could be extended, for instance, to a setting where there is a set of strategic agents with skill $q$ and another set of experts with skill $q_e$.

Furthermore, suppose there are two experts with expertise $q_e$ and $q_f$. Then the equilibrium condition in \Cref{prop:eqmexpert} would be modified to $c^{ef} = V \Phi^{ef}(c^{ef},q,q_e,q_f,n)$, where the expression for $\Phi^{ef}$ would be more involved and $c^{ef}(q_e,q_f)$ is the equilibrium threshold of this game. Observation \eqref{eq:ce1} would now read: for small $q_e$ and $q_f$, $c^*(n+1) < c^{ef}(q_e,q_f) < c^*(n)$. Observation \eqref{eq:ce2} would then read: $c^{ef}(q,q) < c^*(n+2) < c^*(n)$. These lead to the same conclusion that there are critical values $\hat{q}_e$ and $\hat{q}_f$, such that the new threshold is the same as the threshold in a game with $n+2$ (strategic) agents. With two experts, \Cref{ex:probn} and \Cref{ex:probn2} can also be applied to illustrate the same point, i.e. that adding experts can decrease the probability of success.

%%%%%%%%%%%%%%%%%%%%%%%%%%%%%%%%%%%%%%%%%%%%%%%%%%%%%%%%%%%%%%%%%%%%%%
%%%%%%%%%%%%%%%%%%%%%%%%%%%%%%%%%%%%%%%%%%%%%%%%%%%%%%%%%%%%%%%%%%%%%%

\subsection{Multiple Bugs}

The model in this paper can be straightforwardly extended to allow for multiple types and multiplicity of bugs. Suppose there are $L$ types of bugs, indexed by $l \in \{1,\dots,L\}$. The number of type $l$ bugs is a discrete random variable $R^l$ distributed on $\{0,1,\dots\}$ with finite expectation.\footnote{Note that $R^l = 0$ means that the bug does not exist. In practice, the support of $R^l$ is finite as the maximum number of bugs could be taken to be, for instance, the number of characters in the code.} Each bug of type $l$ can be found with probability $q^l$ independently by an agent and yields reward $V^l$ per bug. The (interior and symmetric) equilibrium threshold $c^* = c^*(\boldsymbol{V}, \boldsymbol{R},\boldsymbol{q},n)$ solves:
\begin{equation} \label{eq:multiplebugs}
    c^* = \sum_{l=1}^L V^l \mathbb{E}[R^l] \Phi(c^*, q^l,n).
\end{equation}
Many properties of equilibrium carry over from the one bug case since the RHS of \eqref{eq:multiplebugs} is a linear combination of the function $\Phi$'s. The designer's objective would then be to maximize
\begin{equation}
    \sum_{l=1}^L (W^l-V^l)\mathbb{E}[R^l] P(c^*,q^l,n).
\end{equation}

%%%%%%%%%%%%%%%%%%%%%%%%%%%%%%%%%%%%%%%%%%%%%%%%%%%%%%%%%%%%%%%%%%%%%%
%%%%%%%%%%%%%%%%%%%%%%%%%%%%%%%%%%%%%%%%%%%%%%%%%%%%%%%%%%%%%%%%%%%%%%

%%%%%%%%%%%%%%%%%%%%%%%%%%%%%%%%%%%%%%%%%%%%%%%%%%%%%%%%%%%%%%%%%%%%%%
%%%%%%%%%%%%%%%%%%%%%%%%%%%%%%%%%%%%%%%%%%%%%%%%%%%%%%%%%%%%%%%%%%%%%%

\section{Conclusion} \label{sec:conclusion}

As the empirical literature suggests, bug bounty programs can make an important contribution to the security of businesses and public infrastructures, and private firms. We have provided a simple model to study important dimensions along which such programs can be designed. Of course, numerous further directions can be pursued. For instance, one might introduce entry checks regarding the reputation and past achievements of security researchers to build a favorable pool for finding bugs. Alternatively, would the opposite approach (only allowing greenhorns) be beneficial in a bug bounty scheme, as this would motivate many to participate? Also, one could consider a broader menu of rewards, as researchers may be motivated by monetary rewards as well as by reputation gains, which could be documented by success certificates and which would be valuable as an entry ticket for future bug bounty programs. Finally, one could develop further formulas for how prizes for successful bug finding should be determined. 

%%%%%%%%%%%%%%%%%%%%%%%%%%%%%%%%%%%%%%%%%%%%%%%%%%%%%%%%%%%%%%%%%%%%%%
%%%%%%%%%%%%%%%%%%%%%%%%%%%%%%%%%%%%%%%%%%%%%%%%%%%%%%%%%%%%%%%%%%%%%%
%%%%%%%%%%%%%%%%%%%%%%%%%%%%%%%%%%%%%%%%%%%%%%%%%%%%%%%%%%%%%%%%%%%%%%
%%%%%%%%%%%%%%%%%%%%%%%%%%%%%%%%%%%%%%%%%%%%%%%%%%%%%%%%%%%%%%%%%%%%%%

%\clearpage

\singlespacing

% % That's the necessary style for economics publications
\bibliographystyle{apalike}
% % You can add several .bib-file with comma as separator
\bibliography{bug}

%\clearpage

%\linenumbers
%\singlespacing
\onehalfspacing
%\doublespacing

\appendix

\section{Preliminaries}

We begin with some generalizations of the binomial theorem $\sum_{k=0}^n \binom{n}{k} x^ky^{n-k} = (x+y)^n$.

\begin{lemma} \label{lemma:modbinom}
For $n \in \mathbb{N}$ and $x \neq 0$,
\begin{equation}
\sum_{k=0}^n \binom{n}{k} \frac{x^ky^{n-k}}{k+1} = \frac{1}{n+1}\frac{(x+y)^{n+1}-y^{n+1}}{x}.
\end{equation}
\end{lemma}

\begin{proof}[Proof of \Cref{lemma:modbinom}] Note that $\binom{n}{k} \frac{1}{k+1} = \binom{n+1}{k+1}\frac{1}{n+1}$.
Then,
\begin{equation}
\begin{aligned} \sum_{k=0}^n \binom{n}{k} \frac{x^k y^{n-k}}{k+1} &= \frac{1}{n+1} \sum_{k=0}^n \binom{n+1}{k+1} x^k y^{n-k} = \frac{1}{n+1} \frac{1}{x} \sum_{k=0}^{n} \binom{n+1}{k+1} x^{k+1} y^{(n+1)-(k+1)}
\\ &= \frac{1}{n+1} \frac{1}{x} \left[ \sum_{k=0}^{n+1} \binom{n+1}{k} x^{k} y^{n+1-k} - y^{n+1} \right] 
\\ &= \frac{1}{n+1} \frac{(x+y)^{n+1} - y^{n+1}}{x}.
\end{aligned}
\end{equation}
\end{proof}

%%%%%%%%%%%%%%%%%%%%%%%%%%%%%%%%%%%%%%%%%%%%%%%%%%%%%%%%%%%%%%%%%%%%%%
%%%%%%%%%%%%%%%%%%%%%%%%%%%%%%%%%%%%%%%%%%%%%%%%%%%%%%%%%%%%%%%%%%%%%%

\begin{lemma} \label{lemma:modbinom2} For $n \in \mathbb{N}$ and $x \neq 0$,
\begin{equation} \label{eq:modbinom2}
        \sum_{k=0}^n \binom{n}{k} \frac{x^ky^{n-k}}{(k+1)(k+2)} = \frac{1}{(n+1)(n+2)}\frac{(x+y)^{n+2}-[(n+2)x+y]y^{n+1}}{x^2}.
    \end{equation}
\end{lemma}

\begin{proof}[Proof of \Cref{lemma:modbinom2}]
Note that $\binom{n}{k} \frac{1}{(k+1)(k+2)} = \binom{n+2}{k+2} \frac{1}{(n+1)(n+2)}$. Then,
\begin{equation}
    \begin{aligned} \sum_{k=0}^n \binom{n}{k} \frac{x^k y^{n-k}}{(k+1)(k+2)} &= \frac{1}{(n+1)(n+2)} \sum_{k=0}^n \binom{n+2}{k+2} x^k y^{n-k}
\\ &= \frac{1}{(n+1)(n+2)} \frac{1}{x^2} \sum_{k=0}^{n} \binom{n+2}{k+2} x^{k+2} y^{(n+2)-(k+2)}
\\ &= \frac{1}{(n+1)(n+2)} \frac{1}{x^2} \left[ \sum_{k=0}^{n+2} \binom{n+2}{k} x^{k} y^{n+2-k} - (n+2)x y^{n+1} - y^{n+2}  \right] 
\\ &= \frac{1}{(n+1)(n+2)} \frac{(x+y)^{n+2} - [(n+2)x+y]y^{n+1}}{x^2}.
\end{aligned}
\end{equation}
\end{proof}

%%%%%%%%%%%%%%%%%%%%%%%%%%%%%%%%%%%%%%%%%%%%%%%%%%%%%%%%%%%%%%%%%%%%%%
%%%%%%%%%%%%%%%%%%%%%%%%%%%%%%%%%%%%%%%%%%%%%%%%%%%%%%%%%%%%%%%%%%%%%%

\begin{lemma} \label{lemma:modbinom3} For $n \in \mathbb{N}$ and $x \neq 0$,
    \begin{equation} \label{eq:modbinom3}
        \sum_{k=0}^n \binom{n}{k} \frac{x^ky^{n-k}}{k+2} = \frac{1}{(n+1)(n+2)}\frac{[(n+1)x-y](x+y)^{n+1}+y^{n+2}}{x^2}.
    \end{equation}
\end{lemma}

\begin{proof}[Proof of \Cref{lemma:modbinom3}] 
Note that $\binom{n}{k} \frac{1}{k+2} = \binom{n}{k}\frac{1}{(k+1)(k+2)} + \binom{n}{k}\frac{k}{(k+1)(k+2)}$.
Then,
\begin{equation}
\begin{aligned} 
\sum_{k=0}^n \binom{n}{k} \frac{x^k y^{n-k}}{k+2} &= \sum_{k=0}^n \binom{n}{k} \frac{x^k y^{n-k}}{(k+1)(k+2)} + \sum_{k=0}^n \binom{n}{k} \frac{k x^k y^{n-k}}{(k+1)(k+2)}.
\end{aligned}
\end{equation}
By \Cref{lemma:modbinom2}, the first term is 
\[\frac{1}{(n+1)(n+2)} \frac{(x+y)^{n+2} - [(n+2)x+y]y^{n+1}}{x^2}.\] 
For the second term, we have
\begin{equation}
\begin{aligned}
    &\sum_{k=0}^n \binom{n}{k} \frac{k x^k y^{n-k}}{(k+1)(k+2)} = x \sum_{k=1}^n \binom{n}{k} \frac{k x^{k-1} y^{n-k}}{(k+1)(k+2)} = x \frac{\mathrm{d}}{\mathrm{d}x}\left[ \sum_{k=1}^n \binom{n}{k} \frac{x^k y^{n-k}}{(k+1)(k+2)} \right] \\
    &= x \frac{\mathrm{d}}{\mathrm{d}x}\left[\frac{1}{(n+1)(n+2)} \frac{(x+y)^{n+2} - [(n+2)x+y]y^{n+1}}{x^2} - \frac{y^n}{2} \right] \\
    &= x \left[ \frac{[(n+2)(x+y)^{n+1} - (n+2)y^{n+1}]x^2 - 2x\{(x+y)^{n+2} - [(n+2)x+y]y^{n+1}\}}{(n+1)(n+2) x^4}\right] \\
    &= \frac{1}{(n+1)(n+2)}\frac{(n+2)x(x+y)^{n+1} -(n+2)xy^{n+1} - 2(x+y)^{n+2} + 2(n+2)xy^{n+1}+2y^{n+2}}{x^2} \\
    &= \frac{1}{(n+1)(n+2)}\frac{[(n+2)x-2(x+y)](x+y)^{n+1} + (n+2)xy^{n+1}+2y^{n+2}}{x^2} \\
    &= \frac{1}{(n+1)(n+2)}\frac{(nx-2y)(x+y)^{n+1} + [(n+2)x +2y]y^{n+1}}{x^2}.
\end{aligned}
\end{equation}
Combining the two terms yields
\begin{equation}
\begin{aligned}
    \sum_{k=0}^n \binom{n}{k} \frac{x^k y^{n-k}}{k+2} &= \frac{1}{(n+1)(n+2)} \frac{(x+y)^{n+2} - [(n+2)x+y]y^{n+1}}{x^2} \\ &\qquad +\frac{1}{(n+1)(n+2)}\frac{(nx-2y)(x+y)^{n+1} + [(n+2)x +2y]y^{n+1}}{x^2} \\ &=  \frac{1}{(n+1)(n+2)} \frac{[(x+y) + nx -2y)](x+y)^{n+1} + y^{n+2}}{x^2} \\ &= \frac{1}{(n+1)(n+2)} \frac{[(n+1)x-y](x+y)^{n+1} + y^{n+2}}{x^2}.
\end{aligned}
\end{equation}
\end{proof}

For completeness, we include a version of Bernoulli's Inequality which is used repeatedly in the paper.

\begin{lemma}[Bernoulli's Inequality] \label{lemma:bernoulli}
For $0 < x < 1$ and a positive integer $m$,
\[(1-x)^m < \frac{1}{1+mx}.\]
\end{lemma}

\begin{proof}[Proof of \Cref{lemma:bernoulli}]
For $0 < x < 1$, $\sum_{k=0}^\infty x^k = \frac{1}{1-x}$. Taking the $(m-1)$-th derivative of the identity yields
\begin{equation} \label{eq:geometricdiff}
    \sum_{k=0}^\infty k(k-1)\cdots(k-(m-2)) x^{k-(m-1)} = (m-1)! \frac{1}{(1-x)^m}.
\end{equation}
LHS of \eqref{eq:geometricdiff} is $\sum_{k=m-1}^\infty k(k-1)\cdots(k-(m-2))x^{k-(m-1)} = \sum_{k=0}^\infty \frac{(k+m-1)!}{k!}x^{k}$.
Therefore, it follows that
\begin{equation}
    \frac{1}{(1-x)^m} = 1 + mx + \frac{(m+1)m}{2} x^2 + \frac{(m+2)(m+1)m}{6}x^3 + \cdots > 1+mx.
\end{equation}
\end{proof}

%%%%%%%%%%%%%%%%%%%%%%%%%%%%%%%%%%%%%%%%%%%%%%%%%%%%%%%%%%%%%%%%%%%%%%
%%%%%%%%%%%%%%%%%%%%%%%%%%%%%%%%%%%%%%%%%%%%%%%%%%%%%%%%%%%%%%%%%%%%%%

\section{Proofs} \label{sec:proofs}

\begin{proof}[Proof of \Cref{prop:threshold}]
If $\sigma_i^*: [\underline{c},\overline{c}] \rightarrow \{0,1\}$ is an equilibrium strategy for agent $i$, then it is non-increasing. We prove this by contradiction and suppose that there exists a pair of costs $c < c'$ such that $\sigma_i^*(c) = 0$ and $\sigma_i^*(c') = 1$. Then, $c < c'$ implies
\begin{equation}
\begin{aligned} 
0 &= \mathbb{E}[u_i(0,\sigma^*_{-i}(c_{-i}),c_{i}|c_i = c] \\ &\geq \mathbb{E}[u_i(1,\sigma^*_{-i}(c_{-i}),c_i)|c_i = c] \\ &> \mathbb{E}[u_i(1,\sigma^*_{-i}(c_{-i}),c_i)|c_i = c'] \\ &\geq  \mathbb{E}[u_i(0,\sigma^*_{-i}(c_{-i}),c_{i}|c_i = c'] = 0,
\end{aligned}
\end{equation}
where the first and the last inequalities follow from the definition of equilibrium. The strict inequality which follows from the definition of $u_i$ leads to a contradiction. Thus, any equilibrium strategy is a threshold strategy. Thus, for all $i$, $\sigma_i^* = \sigma_{c^*_i}$ for some $c^*_i \in [\underline{c},\overline{c}]$.
\end{proof}

%%%%%%%%%%%%%%%%%%%%%%%%%%%%%%%%%%%%%%%%%%%%%%%%%%%%%%%%%%%%%%%%%%%%%%
%%%%%%%%%%%%%%%%%%%%%%%%%%%%%%%%%%%%%%%%%%%%%%%%%%%%%%%%%%%%%%%%%%%%%%

\begin{proof}[Proof of \Cref{prop:psi}]
The function $\Psi$ is the expectation over the cost distribution of $p(\bolds_{-i})$ given that other players follow some threshold strategies: $s_j = \sigma_{\hat{c}_j}(c_j)$ for all $j \neq i$. Moreover, since $p(\bolds_{-i})$ is a function only of the number of the other agents who search, we define $\tilde{p}(S_{-i}) \equiv p(\bolds_{-i})$. Then,
\begin{equation}
\begin{aligned}
    \Psi(\hat{\boldc}_{-i}) &= \mathbb{E}[p(\sigma_{\hat{c}_1}(c_1),\dots,\sigma_{\hat{c}_{i-1}}(c_{i-1}),\sigma_{\hat{c}_{i+1}}(c_{i+1}),\dots,\sigma_{\hat{c}_n}(c_n))] \\
    &= \mathbb{E}[\tilde{p}(\sigma_{\hat{c}_1}(c_1)+\dots+\sigma_{\hat{c}_{i-1}}(c_{i-1})+\sigma_{\hat{c}_{i+1}}(c_{i+1})+\dots+\sigma_{\hat{c}_n}(c_n))],
\end{aligned}
\end{equation}
where the expectation is taken over the $c_j$'s. Property (i) follows since the $c_j$'s are distributed identically and independently. 

For property (ii), note that $\tilde{p}(S_{-i})$ is strictly decreasing. This can be shown by applying \Cref{lemma:modbinom} and taking the derivative with respect to $S_{-i}$. Now, fix $j \neq i$ and consider some $\hat{c}'_j > \hat{c}_j$. Define two random variables
\[\Sigma \equiv \sigma_{\hat{c}_1}(c_1)+\dots+\sigma_{\hat{c}_j}(c_j)+\dots+\sigma_{\hat{c}_n}(c_n) \quad \text{and} \quad \Sigma'\equiv \sigma_{\hat{c}_1}(c_1)+\dots+\sigma_{\hat{c}'_j}(c_j)+\dots+\sigma_{\hat{c}_n}(c_n). \]
We claim that $\Sigma'$ first-order stochastically dominates $\Sigma$. For any $x$, we have that $\{ \Sigma \geq x \} \subset \{ \Sigma + \sigma_{\hat{c}'_j} - \sigma_{\hat{c}_j}  \geq x \} = \{ \Sigma' \geq x \}$, where the first inclusion follows because $\sigma_{\hat{c}'_j}(c_j) - \sigma_{\hat{c}_j}(c_j) \geq 0$ for all $c_j$. Thus, for all $x$, $\Pr(\Sigma'\geq x) \geq \Pr(\Sigma \geq x)$ as claimed.

It then follows that $\tilde{p}(\Sigma)$ first-order stochastically dominates $\tilde{p}(\Sigma')$ since $\tilde{p}$ is strictly decreasing. By stochastic dominance, we have
\[\Psi(\hat{c}_1,\dots,\hat{c}'_j,\dots, \hat{c}_n) = \mathbb{E}[\tilde{p}(\Sigma')] < \mathbb{E}[\tilde{p}(\Sigma)] = \Psi(\hat{\boldc}_{-i}),\]
which completes the proof of (ii).

Property (iii) holds because $F(\underline{c}) = 0$ and only the first term ($K=\varnothing$) in the sum survives: $\Psi(\underline{c},\dots,\underline{c}) = q$. On the other hand, $F(\overline{c}) = 1$ and only the last term ($K = N\setminus\{i\}$) in the sum survives: $\Psi(\overline{c},\dots,\overline{c}) = q \sum_{t=0}^{n-1} \binom{n-1}{t} \frac{q^{t}(1-q)^{n-1-t}}{t+1} = \frac{1-(1-q)^n}{n}$ by \Cref{lemma:modbinom}.
\end{proof}

%%%%%%%%%%%%%%%%%%%%%%%%%%%%%%%%%%%%%%%%%%%%%%%%%%%%%%%%%%%%%%%%%%%%%%
%%%%%%%%%%%%%%%%%%%%%%%%%%%%%%%%%%%%%%%%%%%%%%%%%%%%%%%%%%%%%%%%%%%%%%

\begin{proof}[Proof of \Cref{prop:eqm}]
The proof proceeds in three steps. First, we show that under \Cref{assump:symmetric}, the same equilibrium threshold must be used by all agents. Second, we derive a simple expression that an interior equilibrium threshold must satisfy. Third, we show that the threshold $c^*$ is unique and is interior under \Cref{assump:interior}.

\begin{itemize}[leftmargin=*]

    \item[] \textbf{Step 1}. Without loss of generality, let $\boldc^* = (c^*_1,\dots,c^*_n)$ be the interior equilibrium threshold vector such that $c^*_1 \leq \dots \leq c^*_n$. Now, suppose for the sake of contradiction that $c^*_1 < c^*_n$. By definition, we have
    \begin{equation}
        c^*_1 = V\Psi(c^*_2,\dots,c^*_n) \qquad \text{and} \qquad  c^*_n = V\Psi(c^*_1,\dots,c^*_{n-1}).
    \end{equation}
    Combining gives
    \begin{equation}
        \begin{aligned}
        c^*_n - c^*_1 &= V \left[\Psi(c^*_1,c^*_2,\dots,c^*_{n-1}) - \Psi(c^*_2,\dots,c^*_{n-1},c^*_{n}) \right] \\
                    &= - V \left[\Psi(c^*_2,\dots,c^*_{n-1},c^*_{n}) - \Psi(c^*_1,c^*_2,\dots,c^*_{n-1}) \right] \\
                    &= - V \left[\Psi(c^*_n,c^*_2,\dots,c^*_{n-1}) - \Psi(c^*_1,c^*_2,\dots,c^*_{n-1}) \right],
        \end{aligned}
    \end{equation}
    where the second equality takes out the negative sign from the parentheses and the last equality uses the fact that $\Psi$ is symmetric in its arguments. Rearranging gives
    \begin{equation}
    \frac{\Psi(c^*_n,c^*_2,\dots,c^*_{n-1}) - \Psi(c^*_1,c^*_2,\dots,c^*_{n-1})}{c^*_n - c^*_1} = -\frac{1}{V}.
    \end{equation}
    The left-hand side is the slope of $\Psi$ from $c^*_1$ to $c^*_n$, keeping all other arguments fixed. Since $\Psi$ is continuous, it follows by the Mean Value Theorem that there exists $\tilde{c} \in (c^*_1,c^*_n)$ such that 
    \begin{equation}
        \left. \frac{\partial \Psi(c_1,c^*_2,\dots,c^*_{n-1})}{\partial c_1} \right|_{c_1 = \tilde{c}} = -\frac{1}{V},
    \end{equation} 
    which contradicts \Cref{assump:symmetric}.

    \item[] \textbf{Step 2}. We have established that any equilibrium strategy is a threshold strategy in \Cref{prop:threshold}. With \Cref{assump:symmetric}, the threshold equilibrium vector is of the form $\boldc^* = (c^*,\dots,c^*)$ for some $c^*$ that satisfy
    \begin{equation}
        c^* = V\Psi(c^*,\dots,c^*).
    \end{equation}
   We now simplify the expression for $\Psi(c,\dots,c)$. From \eqref{eq:psi},
\begin{equation} \label{eq:psisym}
\Psi(c,\dots,c) = q \sum_{k=0}^{n-1} \left\{ \binom{n-1}{k} F(c)^k(1-F(c))^{n-1-k} \left[ \sum_{t=0}^k \binom{k}{t} q^t(1-q)^{k-t} \frac{1}{t+1} \right] \right\}.
\end{equation}
By \Cref{lemma:modbinom}, the term in the square bracket is
\begin{equation}\sum_{t=0}^k \binom{k}{t} q^t(1-q)^{k-t} \frac{1}{t+1} = \frac{1}{k+1}\frac{1-(1-q)^{k+1}}{q}.
\end{equation}
Then, for $c > \underline{c}$,
\begin{equation}
\begin{aligned}
    &\Psi(c,\dots,c) = \\ &= \sum_{k=0}^{n-1} \binom{n-1}{k} F(c)^k(1-F(c))^{n-1-k}\frac{1-(1-q)^{k+1}}{k+1} \\ &= \sum_{k=0}^{n-1} \binom{n-1}{k} \frac{F(c)^k(1-F(c))^{n-1-k}}{k+1} - (1-q)\sum_{k=0}^{n-1} \binom{n-1}{k} \frac{(F(c)(1-q))^k(1-F(c))^{n-1-k}}{k+1} \\ &=  \frac{1-(1-F(c))^n}{nF(c)} - (1-q)\left(\frac{(1-qF(c))^n-(1-F(c))^n}{nF(c)(1-q)}\right)
    \\ &= \frac{1-(1-qF(c))^n}{nF(c)} = \frac{P(c)}{nF(c)}.
\end{aligned}
\end{equation}
For $c = \underline{c}$, $F(\underline{c}) = 0$ and the only the first ($k=0$) term in the sum survives: $\Psi(\underline{c},\dots,\underline{c}) = q$. Therefore, indeed $\Psi(c,\dots,c) = \Phi(c,q,n)$ as defined in \eqref{eq:phi}. It follows that if an interior equilibrium threshold exists, then it solves $c^* = V\Phi(c^*,q,n)$.

    \item[] \textbf{Step 3}.  We next show that indeed $c^*$ is interior and unique. First observe that $\Phi(c,q,n)$ is strictly decreasing in $c$:
\begin{equation}
\begin{aligned}
    \frac{\partial \Phi}{\partial c} &= \frac{[-n(1-qF(c))^{n-1}(-qf(c))]nF(c) - nf(c)[1-(1-qF(c))^n]}{(nF(c))^2} \\
    &= \frac{nf(c)(1-qF(c))^{n-1}}{(nF(c))^2} \left[ nqF(c) - \frac{1}{(1-qF(c))^{n-1}} + (1-qF)\right] \\
    &= \frac{f(c)(1-qF(c))^{n-1}}{nF(c)^2}\left[1+(n-1)qF(c) - \frac{1}{(1-qF(c))^{n-1}}\right] < 0.
\end{aligned}
\end{equation}
This follows from Bernoulli's Inequality. Therefore, by continuity, $V\Phi(c,q,n)$ has at most one fixed point $c^*$. We claim that such $c^*$ is interior $c^* \in (\underline{c},\overline{c})$ if 
\begin{equation}
    \underline{c} < V \Phi(\underline{c},q,n) \quad \text{and} \quad V\Phi(\overline{c},q,n) < \overline{c}.
\end{equation}
Because $\Phi(\underline{c},q,n) = q$ and $F(\overline{c}) = 1$, the conditions become
\begin{equation}
    \underline{c} < qV \quad \text{and} \quad V\frac{1-(1-q)^n}{n}  < \overline{c},
\end{equation}
which is stated as \Cref{assump:interior}.
\end{itemize}

This completes the proof.
\end{proof}

%%%%%%%%%%%%%%%%%%%%%%%%%%%%%%%%%%%%%%%%%%%%%%%%%%%%%%%%%%%%%%%%%%%%%%
%%%%%%%%%%%%%%%%%%%%%%%%%%%%%%%%%%%%%%%%%%%%%%%%%%%%%%%%%%%%%%%%%%%%%%

\begin{proof}[Proof of \Cref{prop:compstat}] 
We show that $\Phi(c,q,n)$ has the properties as claimed. For $c$, the proof is from the proof of \Cref{prop:eqm}. For $q$, we obtain directly: $\partial \Phi/ \partial q = (1-qF(c))^{n-1} > 0$. For $n$, we show that if $c > \underline{c}$,
\begin{equation}\label{eq:phin}
    \underbrace{\frac{1-(1-qF(c))^n}{nF(c)}}_{\Phi(c,q,n)} > \underbrace{\frac{1-(1-qF(c))^{n+1}}{(n+1)F(c)}}_{\Phi(c,q,n+1)}.
\end{equation}
Inequality \eqref{eq:phin} simplifies to $(1-qF(c))^n <  (1+nqF(c))^{-1}$ which holds by Bernoulli's Inequality. The comparative statics results on $c^*(V,q,n)$ then follows from the properties of $\Phi$ since $c^*$ is the fixed point of $V\Phi$.
\end{proof}

%%%%%%%%%%%%%%%%%%%%%%%%%%%%%%%%%%%%%%%%%%%%%%%%%%%%%%%%%%%%%%%%%%%%%%
%%%%%%%%%%%%%%%%%%%%%%%%%%%%%%%%%%%%%%%%%%%%%%%%%%%%%%%%%%%%%%%%%%%%%%

\begin{proof}[Proof of \Cref{prop:compstatprob}] $P(c,q,n)$ is increasing in $c$ and $q$. Therefore, $P^*(V,q,n) = P(c^*(V,q,n),q,n)$ increases with $V$ and $q$. The example in the main text shows that $P^*$ may increase or decrease with $n$.
\end{proof}

%%%%%%%%%%%%%%%%%%%%%%%%%%%%%%%%%%%%%%%%%%%%%%%%%%%%%%%%%%%%%%%%%%%%%%
%%%%%%%%%%%%%%%%%%%%%%%%%%%%%%%%%%%%%%%%%%%%%%%%%%%%%%%%%%%%%%%%%%%%%%

\begin{proof}[Proof of \Cref{prop:dpdn}]
We start with the fact that for a function $g(n)$,
\begin{equation}
    \frac{\mathrm{d}}{\mathrm{d} n} (g(n))^n = (g(n))^n\left( \frac{ng'(n)}{g(n)} + \ln{g(n)}\right).
\end{equation} 
To why this holds, let $h(n) \equiv (g(n))^n$. Then, $\ln{h(n)} = n\ln{g(n)}$ and taking derivative with respect to $n$ on both sides yields $\frac{h'(n)}{h(n)} = \frac{n g'(n)}{g(n)}+\ln{g(n)}$, which gives the result after some rearrangements. Applying this fact with $g(n) = 1-qF(c(n))$ yields
\begin{equation} \label{eq:Pn1}
\begin{aligned}
    \frac{\mathrm{d}P_n}{\mathrm{d} n} &= \frac{\mathrm{d}}{\mathrm{d} n} 1 - (1-qF(c_n))^n = - \frac{\mathrm{d}}{\mathrm{d} n} (1-qF(c_n))^n \\ &= (1-qF(c_n))^n \left[ \frac{nqf(c_n)}{1-qF(c_n)} \frac{\mathrm{d}c_n}{\mathrm{d}n} - \ln{(1-qF(c_n))}\right],
    \end{aligned}
\end{equation}
where we have used $c_n = c^*(n)$ and $P_n = P^*(n)$ for ease of notation.

Now, implicit differentiation of the equilibrium condition $c_n n F(c_n) = VP_n$ gives
\begin{equation} \label{eq:Pn2}
    c_nF(c_n)+ n \frac{\mathrm{d}c_n}{\mathrm{d}n}F(c_n) + nc_nf(c_n) \frac{\mathrm{d}c_n}{\mathrm{d}n} = V \frac{\mathrm{d}P_n}{\mathrm{d}n} = \frac{c_n n F(c_n)}{P_n} \frac{\mathrm{d}P_n}{\mathrm{d}n}.
\end{equation}
Combining \eqref{eq:Pn1} and \eqref{eq:Pn2} yields 
\begin{equation} \label{eq:Pn3}
\begin{aligned}
    \frac{\mathrm{d}P_n}{\mathrm{d}n} &= (1-qF(c_n))^n \left[ \frac{nqf(c_n)}{1-qF(c_n)} \left( \frac{\frac{c_n n F(c_n)}{P_n} \frac{\mathrm{d}P_n}{\mathrm{d}n}-c_nF(c_n)}{nF(c_n) + n c_n f(c_n)}\right) - \ln{(1-qF(c_n))}\right] \\ &= (1-qF(c_n))^{n-1}\frac{qf(c_n)c_n n F(c_n)}{F(c_n) + c_nf(c_n)} \frac{1}{P_n} \frac{\mathrm{d}P_n}{\mathrm{d}n} \\ &\quad \quad + (1-qF(c_n))^n \left[ \frac{-qf(c_n)c_nF(c_n)}{(1-qF(c_n))(F(c_n)+c_nf(c_n))} - \ln{(1-qF(c_n))} \right].
\end{aligned}
\end{equation}
Solving \eqref{eq:Pn3} for $\mathrm{d}P_n/\mathrm{d}n$ we have
\begin{equation} \label{eq:Pn4}
    \frac{\mathrm{d}P_n}{\mathrm{d}n} = (1-qF(c_n))^n \frac{\left[\frac{-qf(c_n)c_nF(c_n)}{(1-qF(c_n))(F(c_n)+c_nf(c_n))} - \ln{(1-qF(c_n))}\right]}{\left[1-(1-qF(c_n))^{n-1}\frac{qf(c_n)c_nnF(c_n)}{P_n(F(c_n)+c_nf(c_n))}\right]}.
\end{equation}
The denominator in \eqref{eq:Pn4} is
\begin{equation}
\begin{aligned}
    1-(1-qF(c_n))^{n-1}&\frac{qf(c_n)c_nnF(c_n)}{P_n(F(c_n)+c_nf(c_n))} \\ &= \frac{P_n(F(c_n)+c_nf(c_n)) - (1-qF(c_n))^{n-1}qf(c_n)c_n n F(c_n)}{P_n(F(c_n)+c_nf(c_n))} \\ &= \frac{P_n F(c_n)+ c_nf(c_n)[ P_n - (1-qF(c_n))^{n-1}q n F(c_n)]}{P_n(F(c_n)+c_nf(c_n))}.
    \end{aligned}
\end{equation}
It is non-negative because $P_nF(c_n) \geq 0$ and $P_n - (1-qF(c_n))^{n-1}qnF(c_n) = 1-(1-qF(c_n))^{n-1}(1+(n-1)qF(c_n)) \geq 0$ by Bernoulli's Inequality.

Therefore, we have $\mathrm{d}P_n/\mathrm{d}n \geq 0$ if and only if the numerator term in \eqref{eq:Pn4} is non-negative. That is,
\begin{equation}
    \frac{-qf(c_n)c_nF(c_n)}{(1-qF(c_n))(F(c_n)+c_nf(c_n))} \geq \ln{(1-qF(c_n))}.
\end{equation}
Dividing both sides by $-qF(c_n)$ and rearranging yield the condition stated in the proposition.

\end{proof}

%%%%%%%%%%%%%%%%%%%%%%%%%%%%%%%%%%%%%%%%%%%%%%%%%%%%%%%%%%%%%%%%%%%%%%
%%%%%%%%%%%%%%%%%%%%%%%%%%%%%%%%%%%%%%%%%%%%%%%%%%%%%%%%%%%%%%%%%%%%%%

\begin{proof}[Proof of \Cref{prop:clargen}] The equilibrium condition can be written as $c_n n F(c_n) = V P_n$. Assume that $\lim_{n\rightarrow \infty}c_n$ is not equal to $\underline{c}$. Then, since $c_n$ is decreasing in $n$ because of \Cref{prop:compstat}, we have that there exists some value $c'>\underline{c}$, so that $c_n>c'$ for any $n\in \mathbb{N}$. In this case, $c_n n F(c_n)>c'nF(c')$. This holds because $F$ is an increasing function, and, therefore, $F(c_n)>F(c')$. Note that $F(c')>0$, since $c'>\underline{c}$. Therefore, $\lim_{n\rightarrow \infty}{c_nnF(c_n)}\geq \lim_{n\rightarrow \infty}c'nF(c')=\infty$, which can not be equal to $VP_n$, a contradiction.
\end{proof}

%%%%%%%%%%%%%%%%%%%%%%%%%%%%%%%%%%%%%%%%%%%%%%%%%%%%%%%%%%%%%%%%%%%%%%
%%%%%%%%%%%%%%%%%%%%%%%%%%%%%%%%%%%%%%%%%%%%%%%%%%%%%%%%%%%%%%%%%%%%%%

\begin{proof}[Proof of \Cref{prop:largen}]
\begin{itemize}
\item[(i)] Consider the case $\underline{c} = 0$. The proof is by contradiction. First, assume that $\lim_{n\rightarrow \infty}nF(c_n)\neq \infty$. This implies that there exists $m\in \mathbb{N}$ such that there is an infinite sequence of natural numbers $n_1,n_2,\dots$ so that $n_iF(c_{n_i}) < m$. We have
\begin{equation} \label{eq:largenproof}
    c_{n_i} = V \frac{1-(1-qF(c_{n_i}))^{n_i}}{n_iF(c_{n_i})} \geq Vq\frac{1-e^{-qn_iF(c_{n_i})}}{qn_iF(c_{n_i})} > Vq \frac{1-e^{-qm}}{qm} > 0.
\end{equation}
for any $i \in \mathbb{N}$, which contradicts \Cref{prop:clargen} that $c_n$ converges to $0$. The first equality of \eqref{eq:largenproof} follows from the equilibrium condition. The first inequality follows from the fact that $(1-x)^n \leq e^{-nx}$. The second inequality follows because the function $\frac{1-e^{-x}}{x}$ is strictly decreasing. 

The case $\underline{c} > 0$ is less straightforward. The main subtlety is in whether or not the sequence $nF(c_n)$, which may or may not be monotone, converges. If we assume that it does, say to a constant $\kappa$, then
\begin{equation}
    \frac{1-(1-qF(c_n))^n}{nF(c_n)} = \frac{1-\left[\left(1-\frac{q}{1/F(c_n)}\right)^{\frac{1}{F(c_n)}}\right]^{nF(c_n)}}{nF(c_n)}
\end{equation}
implies that
\begin{equation}
    \frac{1-(1-qF(c_n))^n}{nF(c_n)} \rightarrow \frac{1-e^{-q\kappa}}{\kappa}
\end{equation}
by the fact that $F(c_n) \rightarrow 0$, the definition of $e^x$, and continuity of the function. The equilibrium condition then pins down the value of $\kappa$. Without assuming that $nF(c_n)$ converges the proof is more involved and requires several steps.

\begin{itemize}[leftmargin=*]

    \item[] \textbf{Step 1}. Note that $nF(c_n)$ is bounded because
    \begin{equation}
        nF(c_n) = V\frac{1-(1-qF(c_{n}))^{n}}{c_n} \leq \frac{V}{\underline{c}}.
    \end{equation}
    This means that $\lim_{n \rightarrow \infty} nF(c_n) \neq \infty$.

    Moreover, $nF(c_n)$ does not converge to zero. Otherwise, if $nF(c_n) \rightarrow 0$, then taking the limit and applying L'H\^{o}pital's rule to the right-hand side of
    \begin{equation}
        c_n \geq Vq \frac{1-e^{-qnF(c_n)}}{qnF(c_{n})}
    \end{equation}
    implies $\underline{c} \geq Vq$, which contradicts the interiority of the equilibrium.

    \item[] \textbf{Step 2}. We claim that
    \begin{equation} \label{eq:largenproofmain}
        (1-qF(c_n))^n = e^{-qnF(c_n)} + o(1).
    \end{equation}
    To show why \eqref{eq:largenproofmain} holds, fix $n$ and recall the identities $e^{-nx} = \sum_{k=0}^\infty \frac{(-1)^k (nx)^k}{k!}$ and $(1-x)^n = \sum_{k=0}^\infty \binom{n}{k} (-1)^k x^k$, where $\binom{n}{k} = 0$ for $n < k$.
    
    Applying them to $\xi_n \equiv qF(c_n)$, we have
        \begin{equation}
        \begin{aligned}
            e^{-n\xi_n} - (1-\xi_n)^n &= \sum_{k=0}^\infty \frac{n^k}{k!} (-1)^k\xi_n^k - \sum_{k=0}^\infty \binom{n}{k}(-1)^k\xi_n^k \\
            &= \sum_{k=0}^\infty (-1)^k \left[ \frac{1}{k!} - \frac{1}{n^k}\binom{n}{k} \right](n\xi_n)^k.
                   \end{aligned}
        \end{equation}
    Let $B$ be the bound on $n\xi_n$. Then, by the triangle inequality and the facts that $1/k! - \frac{1}{n^k}\binom{n}{k} \geq 0$ and $(n\xi_n)^k \geq 0$, we have
    \begin{equation} \label{eq:largenproofbound}
        \begin{aligned}
         \left|e^{-n\xi_n} - (1-\xi_n)^n \right| &\leq \sum_{k=0}^\infty \left[ \frac{1}{k!} - \frac{1}{n^k}\binom{n}{k} \right](n\xi_n)^k \\ &\leq \sum_{k=0}^\infty \left[ \frac{1}{k!} - \frac{1}{n^k}\binom{n}{k} \right] B^k.
         \end{aligned}
    \end{equation}
    For each $k$,
    \begin{equation}
        \begin{aligned}
        \frac{1}{n^k} \binom{n}{k} &= \frac{1}{k!}  \frac{n(n-1)\cdots (n-(k-1))}{n^k}  \\ &= \frac{1}{k!}  \frac{n^k + O(n^{k-1})}{n^k} = \frac{1}{k!} + O(n^{-1}).
        \end{aligned}
    \end{equation}
    Therefore, taking $n \rightarrow \infty$ in \eqref{eq:largenproofbound} yields the claim in \eqref{eq:largenproofmain}.

    \item[] \textbf{Step 3}. Define $\kappa \in (0,\infty)$ to be the unique solution to 
    \begin{equation}
        \underline{c} = V \frac{1-e^{-q\kappa}}{\kappa}.
    \end{equation}
    
    We now write
    \begin{equation} \label{eq:largenproof2}
    \begin{aligned}
        c_n - \underline{c} &= V \left[\frac{1-(1-qF(c_n))^n}{nF(c_n)} - \frac{1-e^{-q\kappa}}{\kappa}\right] \\ 
        &= Vq \left[\frac{1-e^{-qnF(c_n)}}{qnF(c_n)} - \frac{1-e^{-q\kappa}}{q\kappa} \right] +  o(1),
    \end{aligned}
    \end{equation}
    where the second equality uses facts proved previously: that $\lim_{n\rightarrow \infty} nF(c_n) \notin \{0,\infty\}$ and that $(1-qF(c_n))^n = e^{-qnF(c_n)} + o(1)$.

    Rearranging \eqref{eq:largenproof2}, we get
    \begin{equation}
        \frac{1}{Vq}(c_n - \underline{c}) = \left[ \frac{1-e^{-qnF(c_n)}}{qnF(c_n)} - \frac{1-e^{-q\kappa}}{q\kappa}   \right] + o(1).
    \end{equation}
    Since LHS goes to zero by \Cref{prop:clargen}, it holds that
    \begin{equation} \label{eq:largenproof3}
        \frac{1-e^{-qnF(c_n)}}{qnF(c_n)} \rightarrow \frac{1-e^{-q\kappa}}{q\kappa}.
    \end{equation}

    Then note that $\zeta(x) = (1-e^{-x})/x$ is strictly monotone and continuous, and thus has a continuous inverse. It follows that $\zeta^{-1}\left(\frac{1-e^{-qnF(c_n)}}{qnF(c_n)}\right) = qnF(c_n) \rightarrow q\kappa = \zeta^{-1}\left( \frac{1-e^{-q\kappa}}{q\kappa}\right)$.
\end{itemize}
This completes the proof.

\item[(ii)] From the proof of (i),
\[P_n  = 1-(1-qF(c_n))^n = 1-e^{-qnF(c_n)} + o(1).\]
The result follows from (i) and continuity.
\end{itemize}
\end{proof}

%%%%%%%%%%%%%%%%%%%%%%%%%%%%%%%%%%%%%%%%%%%%%%%%%%%%%%%%%%%%%%%%%%%%%%
%%%%%%%%%%%%%%%%%%%%%%%%%%%%%%%%%%%%%%%%%%%%%%%%%%%%%%%%%%%%%%%%%%%%%%

\begin{proof}[Proof of \Cref{prop:conv}]
    \begin{itemize}
        \item[(i)] From \Cref{prop:largen}, $P_n$ goes to a strictly positive constant. Write $c_nF(c_n) = V \frac{P_n}{n}$. It follows that $c_nF(c_n) \in O(n^{-1})$ and also that $c_nF(c_n) \in \Omega(n^{-1})$.
        \item[(ii)] If $\underline{c} > 0$, then $c_n \rightarrow \underline{c} > 0$. Then,  $F(c_n) = V\frac{P_n/c_n}{n}$ implies the result since $P_n/c_n$ goes to a strictly positive constant.
    \end{itemize}
\end{proof}

%%%%%%%%%%%%%%%%%%%%%%%%%%%%%%%%%%%%%%%%%%%%%%%%%%%%%%%%%%%%%%%%%%%%%%
%%%%%%%%%%%%%%%%%%%%%%%%%%%%%%%%%%%%%%%%%%%%%%%%%%%%%%%%%%%%%%%%%%%%%%

\begin{proof}[Proof of \Cref{prop:largePn}]
Taking the limit of the left-hand side of \eqref{eq:Pncondition} using L'H\^{o}pital's rule yields
\begin{equation}
\lim_{n \rightarrow \infty} \frac{\ln{(1-qF(c_n))}}{\frac{-qF(c_n)}{1-qF(c_n)}} = \lim_{n \rightarrow \infty} \frac{\frac{-qf(c_n)}{1-qF(c_n)}}{\frac{-qf(c_n)(1-qF(c_n)) -q^2f(c_n)F(c_n)}{(1-qF(c_n))^2}} = \lim_{n \rightarrow \infty} 1-qF(c_n) = 1.
\end{equation}
It follows that if the right-hand side of \eqref{eq:Pncondition} is bounded away from 1, the left-hand side eventually overtakes it for sufficiently large $n$. \Cref{assump:costdistribution} is sufficient for this to hold and the proposition follows.
\end{proof}

%%%%%%%%%%%%%%%%%%%%%%%%%%%%%%%%%%%%%%%%%%%%%%%%%%%%%%%%%%%%%%%%%%%%%%
%%%%%%%%%%%%%%%%%%%%%%%%%%%%%%%%%%%%%%%%%%%%%%%%%%%%%%%%%%%%%%%%%%%%%%

\begin{proof}[Proof of \Cref{prop:eqmexpert}]
The steps of the proof follow that of \Cref{prop:eqm} with two modifications. First, the condition for interiority of $c^e$ is now $\underline{c} < \Phi^e(\underline{c},q,q_e,n)$ and $\Phi^e(\overline{c},q,q_e,n) < \overline{c}$. Second, the equilibrium threshold $c^e$ now solves the following equilibrium condition:
\begin{equation} \label{eq:eqmexpertcond}
    c = V (1-q_e) \Phi(c,q,n) + Vq_e \tilde{\Phi}(c,q,n),
\end{equation}
where
\begin{equation} \label{eq:phitilde}
    \tilde{\Phi}(c,q,n) \equiv  q \sum_{k=0}^{n-1} \binom{n-1}{k} F(c)^k(1-F(c))^{n-1-k} \left[ \sum_{t=0}^k \binom{k}{t} q^t(1-q)^{k-t} \frac{1}{t+2} \right] .
\end{equation}
There are two terms on the right-hand side of \eqref{eq:eqmexpertcond}. First, if the expert does not find the bug (with probability $1-q_e$) then the expected reward for the agents is as before. Second, if the expert finds the bug (with probability $q_e$), then the prize is split in one more additional way\textemdash hence the term $\frac{1}{t+2}$ in the expression for $\tilde{\Phi}$.  We now simplify $\tilde{\Phi}(c,q,n)$ by using the stated lemmata and show that it is strictly decreasing in $c$.

Assume first that $c > \underline{c}$. Using \Cref{lemma:modbinom3},
\begin{equation}
    \sum_{t=0}^k \binom{k}{t} q^t(1-q)^{k-t} \frac{1}{t+2} = \frac{1}{(k+1)(k+2)} \frac{(k+2)q - 1 + (1-q)^{k+2}}{q^2}.
\end{equation}
And hence
\begin{equation}
\begin{aligned}
\tilde{\Phi}(c,q,n) &= \sum_{k=0}^{n-1} \binom{n-1}{k} F(c)^k(1-F(c))^{n-1-k}\frac{1}{k+1} \\ &- \frac{1}{q}\sum_{k=0}^{n-1} \binom{n-1}{k} F(c)^k(1-F(c))^{n-1-k}\frac{1}{(k+1)(k+2)} \\ &+ \frac{(1-q)^2}{q} \sum_{k=0}^{n-1} \binom{n-1}{k} (F(c)(1-q))^k(1-F(c))^{n-1-k}\frac{1}{(k+1)(k+2)}.
\end{aligned}
\end{equation}
Applying \Cref{lemma:modbinom} to the first term and \Cref{lemma:modbinom2} to the second and third terms on the right-hand side yields
\begin{equation}
    \begin{aligned}
    \tilde{\Phi}(c,q,n) &= \frac{1-(1-F(c))^n}{nF(c)} \\ &- \frac{1-(1+nF(c))(1-F(c))^n}{n(n+1)qF(c)^2} \\ &+ \frac{(1-qF(c))^{n+1}-(1+nF(c)-(n+1)qF(c))(1-F(c))^n}{n(n+1)qF(c)^2},
    \end{aligned}
\end{equation}
which simplifies to
\begin{equation}
    \tilde{\Phi}(c,q,n) = \frac{1}{nF(c)} + \frac{(1-qF(c))^{n+1} - 1}{n(n+1)qF(c)^2},
\end{equation}
which can be readily verified to be strictly decreasing in $c$.
The right-hand side of the equilibrium condition is then
\begin{equation}
    V (1-q_e) \Phi(c,q,n) + Vq_e \left[ \frac{1}{nF(c)} + \frac{(1-qF(c))^{n+1} - 1}{n(n+1)qF(c)^2} \right].
\end{equation}
Simplifying gives the expression
\begin{equation}
    V \Phi(c,q,n) - V q_e \frac{1- (1-qF(c))^n(1+nqF(c))}{n(n+1)qF(c)^2}
\end{equation}
which holds true for all $c > \underline{c}$. If $c = \underline{c}$, then $\tilde{\Phi} = q/2$ and $\Phi = q$. The right-hand side of the equilibrium condition is then $V q - V q_e q/2 = Vq(1-q_e/2)$ as defined in \eqref{eq:phiexpert}. The function $\Phi^e$ is strictly decreasing in $c$ since it is a combination of $\Phi$ and $\tilde{\Phi}$ both of which are strictly decreasing in $c$. Existence and uniqueness of a fixed point $c^e$ follow.
\end{proof}

%%%%%%%%%%%%%%%%%%%%%%%%%%%%%%%%%%%%%%%%%%%%%%%%%%%%%%%%%%%%%%%%%%%%%%
%%%%%%%%%%%%%%%%%%%%%%%%%%%%%%%%%%%%%%%%%%%%%%%%%%%%%%%%%%%%%%%%%%%%%%

\begin{proof}[Proof of \Cref{prop:cexpert}]
Using the indifference conditions \eqref{eq:symeqm} and \eqref{eq:eqmexpert}, we have
\begin{equation}     
\Phi^e(c^e(\hat{q}_e),q,n,\hat{q}_e) = \Phi(c^*(n+1),n+1,q).
\end{equation}
The expression for $\hat{q}_e$ follows after some algebra. The rest of the proof is outlined in the main text.
\end{proof}

\begin{proof}[Proof of \Cref{prop:psim}]
Recall that $\Psi^m$ is the expectation of the probability that agent $i$ is the $m$-th agent to find the bug conditioning searching, and on other agents using threshold strategies. That is,
\[\Psi^m(\hat{\boldc}_{-i}) = \mathbb{E}[p^m(\sigma_{\hat{c}_1}(c_1),\dots,\sigma_{\hat{c}_{i-1}}(c_{i-1}),\sigma_{\hat{c}_{i+1}}(c_{i+1}),\dots,\sigma_{\hat{c}_n}(c_n))].\]

We write $\tilde{p}^m(S_{-i}) \equiv p^m(\bolds_{-i})$. Fix $S_{-i}$, then property (i) follows because
\begin{equation}
\begin{aligned}
    \frac{1}{q} \sum_{m=1}^n \tilde{p}^m(S_{-i}) &= \sum_{m=1}^{S_{-i}+1} \sum_{t=m-1}^{S_{-i}} \binom{S_{-i}}{t} q^t(1-q)^{S_{-i} - t}\frac{1}{t+1} \\ &= \sum_{t=0}^{S_{-i}} \sum_{m=1}^{t+1} \binom{S_{-i}}{t} q^t(1-q)^{S_{-i} - t}\frac{1}{t+1}   \\ &= \sum_{t=0}^{S_{-i}} (t+1) \binom{S_{-i}}{t} q^t(1-q)^{S_{-i} - t}\frac{1}{t+1}  \\ &= \sum_{t=0}^{S_{-i}} \binom{S_{-i}}{t} q^t(1-q)^{S_{-i} - t} \\ &= 1.
\end{aligned}
\end{equation}

For property (ii), note that
\begin{equation}
\tilde{p}^m(S_{-i}) - \tilde{p}^{m+1}(S_{-i})  = \left\{ \begin{array}{lcl}
      0 & \mbox{if} & S_{-i} < m-1 \\ 
    q \binom{S_{-i}}{m-1} q^{m-1}(1-q)^{S_{-i} - (m-1)}\frac{1}{m} & \mbox{if} & S_{-i} \geq m-1, \\
\end{array}\right.
\end{equation}
which is non-negative for all $S_{-i}$ and strictly positive for some $S_{-i}$. Replacing $S_{-i}$ with $\sum_{j\neq i} \sigma_{\hat{c}_j}(c_j)$ in the above difference and taking the expectation over the $c_j$'s imply property (ii).

For property (iii), $\Psi^1$ is strictly decreasing in $c_j$ from \Cref{prop:psi}. It suffices to show that for $m \neq 1$, $\Psi^m$ can increase in $c_j$. This is true by property (v). Lastly, property (iv) is from \Cref{prop:psi} and property (v) follows by inspection.
\end{proof}

\begin{proof}[Proof of \Cref{prop:eqmmult}]
The steps of the proof follow that of \Cref{prop:eqm} with appropriate modifications for the conditions for uniqueness and interiority. 
\end{proof}

\begin{proof}[Proof of \Cref{prop:optimalprize}]
To find the highest and lowest fixed points of $\sum_{m=1}^n v^m \Phi^m(\hat{c})$ over $\boldv \in \mathcal{V}$, we simply solve the following two linear programs for a fix $\hat{c}$:
\[\max_{\boldv \in \mathbb{R}^n} v^1 \Phi^1 + \cdots + v^n \Phi^m
\st v^1 \geq \cdots \geq v^n \geq 0 \quad \text{and} \quad v^1+\cdots+v^n = V\]
and
\[\min_{\boldv \in \mathbb{R}^n} v^1 \Phi^1 + \cdots + v^n \Phi^m
\st v^1 \geq \cdots \geq v^n \geq 0 \quad \text{and} \quad v^1+\cdots+v^n = V\]
Now since $\Phi^m$ is a ``slice'' of $\Psi^m$ along the ``diagonal'', \Cref{prop:psim} implies that for all $\hat{c}$, $\Phi^1(\hat{c}) > \Phi^2(\hat{c}) > \dots > \Phi^n(\hat{c})$ and that $\sum_{m=1}^n \Phi^m(\hat{c}) = q$. It follows that $(V,0,\dots,0)$ solves the first linear program, while $(V/n,\dots,V/n)$ solves the second. \Cref{prop:eqmmult} then implies the statements on maximizing and minimizing the probability of success.
\end{proof}

%%%%%%%%%%%%%%%%%%%%%%%%%%%%%%%%%%%%%%%%%%%%%%%%%%%%%%%%%%%%%%%%%%%%%%
%%%%%%%%%%%%%%%%%%%%%%%%%%%%%%%%%%%%%%%%%%%%%%%%%%%%%%%%%%%%%%%%%%%%%%
%%%%%%%%%%%%%%%%%%%%%%%%%%%%%%%%%%%%%%%%%%%%%%%%%%%%%%%%%%%%%%%%%%%%%%

%%%%%%%%%%%%%%%%%%%%%%%%%%%%%%%%%%%%%%%%%%%%%%%%%%%%%%%%%%%%%%%%%%%%%%
%%%%%%%%%%%%%%%%%%%%%%%%%%%%%%%%%%%%%%%%%%%%%%%%%%%%%%%%%%%%%%%%%%%%%%
%%%%%%%%%%%%%%%%%%%%%%%%%%%%%%%%%%%%%%%%%%%%%%%%%%%%%%%%%%%%%%%%%%%%%%
%%%%%%%%%%%%%%%%%%%%%%%%%%%%%%%%%%%%%%%%%%%%%%%%%%%%%%%%%%%%%%%%%%%%%%

\end{document}